%% file: main.tex
\newcommand{\dd}{\mathrm{d}}

\documentclass[superscriptaddress, nofootinbib, amsmath,amssymb, aps, prd, twocolumn]{revtex4-2}

\usepackage{comment}
\usepackage{graphicx}
\usepackage{dcolumn}
\usepackage{bm}
\usepackage{hyperref}
\usepackage{xcolor}
\usepackage{acronym}
\usepackage{lipsum}

\usepackage[normalem]{ulem}

\input{acronyms.tex}

\begin{document}


\title{Mapping of Spacetimes with Neural Networks} 
    
\author{Estuti Shukla}
\email{estuti@psu.edu}
\affiliation{Institute for Gravitation and the Cosmos, The Pennsylvania State University, University Park PA 16802, USA}
\affiliation{Department of Physics, The Pennsylvania State University, University Park PA 16802, USA}
\author{David Radice}
\affiliation{Institute for Gravitation and the Cosmos, The Pennsylvania State University, University Park PA 16802, USA}
\affiliation{Department of Physics, The Pennsylvania State University, University Park PA 16802, USA}
\affiliation{Department of Astronomy \& Astrophysics, The Pennsylvania State University, University Park PA 16802, USA}
\author{Rossella Gamba}
\affiliation{Department of Physics, University of California, Berkeley, CA 94720, USA}
\author{Sebastiano Bernuzzi}
\affiliation{Theoretisch-Physikalisches Institut, Friedrich-Schiller-Universit\"at Jena, 07743, Jena, Germany}
\date{\today}

\begin{abstract}
Gauge freedom in numerical relativity complicates
the interpretation of simulation coordinates and
the comparison of different descriptions of the
same spacetime. We present a data-driven neural network method for learning coordinate maps from numerical metric data to a prescribed target gauge. The network is trained by matching the transformed source metric to the target metric, without assuming a functional form for the coordinate transformation. We first validate the method on analytical Schwarzschild and Kerr geometries, recovering known transformations up to the isometries of the target spacetime. We then apply it to numerical black hole evolutions in moving-puncture coordinates. For a non-spinning black hole, the learned map relates the time-dependent numerical metric to a stationary Schwarzschild metric in ingoing Eddington--Finkelstein coordinates. For a spinning puncture, the method provides an approximate identification with a Kerr background
despite departures from Kerr in the numerical
spacetime. We use this identification to construct curvature perturbations and evolve them with the Teukolsky equation, obtaining waveforms that closely match those extracted from the full nonlinear simulation. These results demonstrate that learned coordinate maps can relate numerical spacetimes to analytical backgrounds, with potential applications to cross-code validation and black hole perturbation theory.
\end{abstract}                 
\maketitle

\section{Introduction}
\Ac{gw} astronomy has entered the era of precision science, with nearly 400 \ac{bbh}, \ac{bns} and \ac{bhns} events cataloged by the \ac{lvk} collaboration~\cite{LIGOScientific:2014pky, VIRGO:2014yos, KAGRA:2020tym} at the end of O4b~\cite{LIGOScientific:2026wfs}. Next-generation (XG) \ac{gw} detectors, including \ac{et}~\cite{Punturo:2010zz,ET:2019dnz}, \ac{ce}~\cite{Reitze:2019iox,Evans:2021gyd} and the space-based detectors like \ac{lisa}~\cite{LISA:2017pwj}, will place increasingly stringent demands on waveform accuracy. We expect that for sufficiently loud sources waveform-model biases can exceed statistical uncertainties~\cite{Purrer:2019jcp,Ferguson:2020xnm, Jan:2023raq, Kapil:2024zdn}. Achieving higher accuracy requires both more accurate \ac{nr} simulations~\cite{Rashti:2024yoc, Shukla:2025kvc} and the ability to rigorously cross-validate waveforms produced by various \ac{nr} codes~\cite{Hannam:2009hh,Habib:2025bkb,Hinder:2013oqa}. Given that \ac{nr} codes differ in their choice of how Einstein's field equations are cast, discretized, gauge-fixed, and supplied with initial data, agreement between codes is a non-trivial test. Cross-code validation serves as an important tool to verify that two independent codes converge to the same physical solution and help identify discrepancies that may not be apparent from convergence studies within a single code. Accurate, independently validated \ac{nr} waveforms support the calibration and validation of \ac{eob} models~\cite{Pompili:2023tna,Albanesi:2025txj,vandeMeent:2023ols,Nagar:2024oyk,Gamba:2024cvy,Ramos-Buades:2021adz,Khalil:2021txt,Gamboa:2024hli,Gamboa:2024imd,Gamboa:2026jht,Pan:2013rra,Babak:2016tgq,Ossokine:2020kjp,Khalil:2023kep,Ramos-Buades:2023ehm,Estelles:2025zah,Akcay:2020qrj,Gamba:2021ydi}, phenomenological models~\cite{Pratten:2020ceb,Estelles:2021gvs,Hamilton:2021pkf,Planas:2025feq} as well as construction of \ac{nr}-based surrogate models~\cite{Varma:2018mmi,Varma:2019csw,Blackman:2017pcm,Islam:2022laz,Rink:2024swg,GramaxoFreitas:2024bpk,Purrer:2026wgp} used in \ac{gw} inference.

Across the \ac{nr} codes used in the community, two gauge families dominate: the moving puncture gauge~\cite{Campanelli:2005dd,Baker:2005vv}, employed in codes like \texttt{AthenaK}~\cite{Zhu:2024utz,Fields:2024pob}, \texttt{GR-Athena++}~\cite{Daszuta:2021ecf}, \texttt{Einstein Toolkit}~\cite{Loffler:2011ay}, \texttt{GRChombo}~\cite{Clough:2015sqa}, and \texttt{BAM}~\cite{Bruegmann:2006ulg}, and the \ac{ghg}~\cite{Pretorius:2005gq,Pretorius:2004jg}, used in \texttt{SpEC}~\cite{Scheel:2008rj} and \texttt{SpECTRE}~\cite{Lovelace:2024wra}. When evolving equivalent physical initial conditions, two codes with different gauge conditions can assign different coordinates
to corresponding spacetime points, making direct comparison of the raw bulk metric data highly non-trivial. 

Cross-code validation therefore typically relies on quantities extracted at future null infinity $\mathcal{I}^{+}$ such as the Weyl scalar $\psi_4$ or the \ac{gw} strain obtained via extrapolation~\cite{Boyle:2009vi,Taylor:2013zia} or Cauchy-characteristic extraction (CCE)~\cite{Reisswig:2009us,Moxon:2021gbv,Babiuc:2010ze}. However, even these asymptotic quantities carry a residual freedom under the \ac{bms} transformations that must be fixed before a consistent comparison can be made~\cite{Mitman:2021xkq,Mitman:2022kwt}. Agreement of the asymptotic waveforms does not identify corresponding spacetime points in simulations with different gauges. Establishing this correspondence requires an explicit coordinate map in the bulk spacetime.

In this work, we develop a \ac{nn} framework for learning coordinate maps directly from numerical metric data, with the aim of enabling point-wise comparisons of the bulk spacetime between codes evolving the same physical system. Such a comparison could provide a complementary diagnostic to asymptotic methods, helping to isolate where in the spacetime two codes begin to disagree, and offering deeper insight into which region of the physical spacetime each \ac{nr} gauge is effectively probing. The key observation is that two physically equivalent spacetimes differing only in gauge are related by a diffeomorphism, which we propose to learn directly from the metric data using \ac{nn}. Our data-driven approach learns this coordinate map by matching the transformed numerical metric to a prescribed target metric, without assuming a functional form for the gauge transformation. Beyond cross-code validation, this approach has a natural application in \ac{bh} perturbation theory: schemes such as the Lazarus project~\cite{Baker:2001sf,Campanelli:2005ia} require an identification of the numerical spacetime with a Kerr background, together with a suitable tetrad, to construct initial data for $\Psi_4$ and its time derivative for evolution with the Teukolsky equation. In the Lazarus approach, this identification used curvature invariants and coordinate prescriptions tailored to the near-Kerr regime. 
 
In spherical symmetry, the spacetime of a numerical puncture evolution has been related to the Schwarzschild geometry in several works~\cite{Hannam:2006vv,PhysRevD.75.067502,Brown:2007tb,Hannam:2008sg,2011PhRvD..83f4022T}. Beyond spherical symmetry, efforts to interpret simulation data in the strong-field region have mostly relied on coordinate-independent curvature quantities. Zhang et al.~\cite{Zhang:2012ky} built radial and polar coordinates from curvature in \ac{nr} simulations, and Sadiq et al.~\cite{Sadiq:2018eyd} compared analytical and numerical spacetimes locally using geodesic deviation. Another related approach to learning coordinate transformations with \acp{nn} was developed by Liu~\&~Tegmark~\cite{Liu:2021azq}. Their method searches for coordinates that make a chosen
symmetry apparent, using a loss that measures how closely
the transformed system satisfies that symmetry. In their Schwarzschild example, they recovered Gullstrand--Painlev\'e coordinates, whose spatial metric is flat. In shape analysis, Celledoni et al.~\cite{Celledoni_2023} developed neural architectures that represent diffeomorphisms and optimize the alignment of curves and surfaces. In our previous work~\cite{Shukla:2026fhq}, a learned map to \ac{ks}
coordinates allowed us to identify the geometry reconstructed
by a physics-informed neural network as Schwarzschild
spacetime. Here, we develop this method further for
numerical metric data.

We first demonstrate that \acp{nn} can recover known coordinate transformations for analytical Schwarzschild and Kerr metrics, up to isometries of the target spacetime. We then apply the
method to numerical simulations of a spherically symmetric
non-spinning \ac{bh} and a perturbed spinning \ac{bh}. For the
non-spinning case, the learned map brings the time-dependent
moving-puncture metric into ingoing \ac{ef} form. For the spinning case, we jointly fit the coordinate map and the mass and spin of a Kerr background in \ac{ks} coordinates. We use this map to construct initial data for Teukolsky evolution and compare the resulting waveforms with those extracted directly from the \ac{nr} simulation.

The structure of the paper is as follows. In Sec.~\ref{sec:bg}, we review gauge freedom in \ac{nr}, formulate the coordinate
mapping problem, and describe our \ac{nn} architecture
and training procedure. In Sec.~\ref{sec:setup}, we present
the numerical simulations used to generate the metric data
and describe the construction and evolution of Teukolsky
initial data. In Sec.~\ref{sec:results}, we present analytical tests, applications to non-spinning and spinning puncture simulations, and comparisons between perturbative and numerical waveforms. We conclude with a summary of our
findings and the limitations of the method in Sec.~\ref{sec:summary}.

We use geometric units with $G=c=1$, unless otherwise stated.

\section{Background}
\label{sec:bg}
The theory of \ac{gr} assumes that spacetime is a four-dimensional
Lorentzian manifold $(\mathcal{M},g)$. On a coordinate patch $U\subset\mathcal{M}$,
the spacetime metric $g$ is represented by sixteen component functions
$g_{\alpha\beta}(x)$ of the four coordinates $(x^0,x^1,x^2,x^3)$, of which
ten are independent because $g_{\alpha\beta}=g_{\beta\alpha}$. Einstein's
equations supply ten equations for these ten unknowns, but the equations
are not independent: the contracted Bianchi identity,
$\nabla_\mu G^{\mu\nu}=0$, holds for every metric and imposes four
differential relations among them. The field equations therefore
determine the metric only up to four arbitrary functions, which
correspond to the freedom to choose the four spacetime coordinates.
This freedom is known as the gauge freedom of
\ac{gr}~\cite{Gourgoulhon:2007ue,Carroll:1997ar}.

\begin{figure}
 \centering
  \includegraphics[width=\columnwidth]{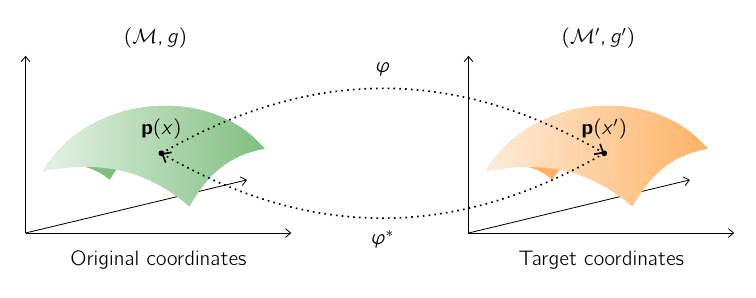}
 \caption{The diffeomorphism $\varphi$ maps the manifold $(\mathcal{M}, g)$ in original coordinates to $(\mathcal{M}', g')$ in target coordinates, with the pullback $\varphi^* g'$ relating the two metric representations via Eq.~\eqref{eq:metric_transform}.}
 \label{fig:manifold}
 \end{figure}

General covariance ensures that Einstein's equations
retain their tensorial form under changes of coordinates.
The equivalence of different descriptions of the same
geometry can also be expressed through diffeomorphisms,
which are smooth bijective maps with smooth
inverses. For the spacetimes considered in Fig.~\ref{fig:manifold}, two descriptions $(\mathcal{M},g)$ and $(\mathcal{M'},g')$ represent the same geometry if there exists a diffeomorphism $\varphi:\mathcal{M}\rightarrow\mathcal{M}'$ satisfying $\varphi^{*}g'=g$, where $\varphi^{*}g'$ denotes the pullback of $g'$ under $\varphi$. In local coordinates $\{x^{\mu}\}$ on $U\subset\mathcal{M}$ and
$\{x'^{\alpha}\}$ on $\varphi(U)\subset\mathcal{M}'$, the pullback of the
metric satisfies the following condition,
\begin{equation}
    \label{eq:metric_transform}
    g_{\mu\nu}(x) = g'_{\alpha\beta}\!\left(x'\right)
    \frac{\partial x'^{\alpha}}{\partial x^{\mu}}
    \frac{\partial x'^{\beta}}{\partial x^{\nu}},
\end{equation}
where $x' = \varphi(x)$.

While the diffeomorphism invariance of \ac{gr}
ensures that the underlying physics is independent
of coordinate choice, it introduces a practical
complication: two representations of the same spacetime
do not, by themselves, specify how their points should
be identified. This poses an obstacle to point-wise
comparisons of numerical solutions produced by different
\ac{nr} codes which may evolve the same physical system
in different gauges, as discussed in the next subsection.

\subsection{Gauge in Numerical Relativity}
 Ill-suited choices for gauge conditions can lead to coordinate singularities, grid distortions and numerical instabilities~\cite{Gourgoulhon:2007ue}. Two approaches have proven robust and are widely used: the moving-puncture gauge and the \ac{ghg}. We briefly describe each in turn.

\subsubsection{Moving Puncture Gauge}
\label{sec:bg.punc}
To solve Einstein's field equations numerically, one commonly uses a Cauchy formulation, in which initial data are specified on a spacelike hypersurface and evolved forward in time~\cite{Foures-Bruhat:1952grw}. In the $3+1$ formalism~\cite{Gourgoulhon:2007ue}, spacetime is foliated into spacelike hypersurfaces $\Sigma_t$, labelled by a time coordinate $t$. The \ac{adm} formulation~\cite{Arnowitt:1959ah} expresses Einstein's equations as evolution equations for the spatial geometry, together with the Hamiltonian and momentum constraints. Obtaining a well-posed evolution system additionally requires suitable choices of gauge conditions.

In the $3+1$ decomposition, the spacetime metric is written as
\begin{equation}
    \label{eq: ADMmetric}
    \dd s^2 = -\alpha^2\,\dd t^2
    + \gamma_{ij}
    \left(\dd x^i+\beta^i\,\dd t\right)
    \left(\dd x^j+\beta^j\,\dd t\right),
\end{equation}
where $\alpha$ is the lapse function, $\beta^i$ is the shift vector, and $\gamma_{ij}$ is the induced metric on $\Sigma_t$. The lapse relates coordinate-time
intervals to the proper time measured by observers normal to the hypersurfaces, while the shift specifies how spatial coordinates are carried between successive
slices. Once the initial coordinates are chosen, prescriptions for the lapse and shift determine their subsequent evolution ~\cite{Gourgoulhon:2007ue}.

Weak hyperbolicity limits the standard \ac{adm} system for commonly adopted gauge prescriptions, motivating the development of alternative evolution formulations. The BSSNOK reformulation~\cite{Nakamura:1987zz,Shibata:1995we,Baumgarte:1998te} addressed this by introducing a conformally rescaled spatial metric, evolving the trace $K$ of the extrinsic curvature independently of its conformally rescaled trace-free part $\tilde{A}_{ij}$, and promoting the conformal
connection functions $\tilde{\Gamma}^i$ to evolved variables.
Together with appropriate gauge conditions these changes render the system strongly hyperbolic~\cite{Sarbach:2002bt,Beyer:2004sv} and suitable for long-term evolution.

The Z4 formulation~\cite{Bona:2003fj} extends Einstein's equations by introducing additional fields associated with constraint violations and allows them to propagate through the numerical grid~\cite{Bernuzzi:2009ex}. Constraint-damping terms can additionally suppress
violations during the evolution~\cite{Gundlach:2005eh}. Bernuzzi and Hilditch~\cite{Bernuzzi:2009ex} introduced the conformal Z4c formulation, which was
subsequently applied to compact-binary simulations by Hilditch et al.~\cite{Hilditch:2012fp}. Z4c is used in \texttt{GR-Athena++}~\cite{Daszuta:2021ecf}
and \texttt{AthenaK}~\cite{Zhu:2024utz}.
A related conformal formulation, CCZ4~\cite{Alic:2011gg}, is available in codes such as \texttt{GRChombo}~\cite{Clough:2015sqa}. In this work, we use a spherically symmetric Z4c solver \texttt{GRHyps}~\cite{Bernuzzi:2009ex} and \texttt{AthenaK} to generate the numerical metric data, as described in Sec.~\ref{sec:setup}.

A particularly robust and widely adopted family of gauge conditions within the BSSNOK and conformal Z4 approach is the moving puncture gauge~\cite{Campanelli:2005dd,Baker:2005vv}, which uses the \textit{$1+\log$} slicing~\cite{Bona:1994dr} condition for the lapse,
\begin{equation}
    \partial_t \alpha  = -2\alpha K + \beta^i \partial_i \alpha,
\end{equation}
where $K$ is the trace of the extrinsic curvature. This slicing condition helps prevent the hypersurfaces from reaching the physical singularity by driving the lapse $\alpha$ toward zero in strong-field region. However, this can cause grid stretching which is controlled by the $\tilde{\Gamma}$-driver shift condition~\cite{Alcubierre:2002kk},
\begin{align}
    \partial_t \beta^i =  \tilde{\Gamma}^i 
         + \beta^j \partial_j \beta^i 
         - \eta\, \beta^i,
\end{align}
where $\tilde{\Gamma}^i$ is a dynamical variable of the Z4c formulation and $\eta$ is a damping parameter.

Standard single-puncture wormhole initial data describe a spatial slice with two asymptotically flat ends. The second end is mapped to the coordinate origin, which is referred to as the
puncture~\cite{Brandt:1997tf}. It therefore corresponds to an asymptotic region of the initial slice, rather than the black hole's curvature singularity. In the continuum limit, the evolved slices keep extending to this second asymptotic end, and the data never settle into a globally stationary state. In practice, however, the $\tilde{\Gamma}$-driver shift carries the innermost grid points across the horizon. The portion of the slice extending towards the second asymptotic end is then squeezed between the puncture and the nearest grid point, where the code can no longer resolve it. This acts as a natural excision~\cite{Brown:2007tb,Hannam:2008sg}. The resolved part of each slice can then approach the stationary trumpet slice of the $1+\log$ condition as the simulation evolves.

The moving puncture gauge has proven remarkably successful across a broad range of binary configurations, including \ac{bbh}, \ac{bns} and \ac{bhns} systems. In this work, we focus on black-hole evolutions using the Z4c formulation with moving puncture gauge.

\subsubsection{Generalised Harmonic Gauge}

In the \ac{gh} formulation~\cite{Pretorius:2004jg,Lindblom:2005qh}, the gauge is
imposed directly on the spacetime coordinates rather than through separate
evolution equations for the lapse and shift. Each coordinate $x^{\mu}$ is
treated as a scalar field obeying a wave equation,
\begin{equation}
    \Box\, x^{\mu} = H^{\mu}(x, g),
\end{equation}
where $\Box \equiv g^{\mu\nu}\nabla_{\mu}\nabla_{\nu}$ and the gauge source
functions $H^{\mu}$ may be chosen freely. Setting $H^{\mu}=0$ recovers ordinary
harmonic coordinates, while non-zero sources provide the flexibility needed to
keep the coordinates well behaved in dynamical, strong-field
spacetimes~\cite{Pretorius:2004jg}. Using this approach, Pretorius~\cite{Pretorius:2005gq} obtained the first
complete simulation of a \ac{bbh} inspiral, merger and ringdown. The \acs{gh} formulation underlies the
\texttt{SpEC} code~\cite{Lindblom:2005qh,Scheel:2008rj} and its discontinuous-Galerkin successor \texttt{SpECTRE}~\cite{Lovelace:2024wra}.

\subsection{Learning the Gauge Transformation}
\label{sec:bg.gauge}
The preceding discussion establishes that a single physical spacetime can be represented in different coordinate descriptions. In this work, we address the following problem: given numerical metric data $g_{\mu\nu}(x_{\mathrm{NR}})$ in some \ac{nr} gauge, we \emph{prescribe} the functional form of the target spacetime metric in the target frame, $g'_{\alpha\beta}(x')$ , such as the Schwarzschild
metric in spherical-polar coordinates or the Kerr
metric in \ac{ks} coordinates. We seek a coordinate map $\varphi_{\rm NN}: x_{\mathrm{NR}} \mapsto x'$, parametrized with a \ac{nn}. Selected parameters of the target metric, such as its mass and spin, may also be determined during the fit. We use the Jacobian of the learned coordinate map
to transform the numerical metric components according
to Eq.~\eqref{eq:metric_transform}. The network is then
trained by minimizing the difference between these
transformed components and the target metric components
evaluated at the corresponding mapped points.

We distinguish two cases. In the first case, the \ac{nr} simulation and the
target metric describe the same spacetime geometry
in different coordinates. An exact coordinate transformation therefore exists for the underlying
solution. In practice, numerical errors in the simulation and inaccuracies in the fitted map
can leave a residual metric mismatch. In the second case, the \ac{nr} spacetime contains a physical departure from the background family, so no coordinate transformation and choice of background parameters can yield an exact match. We write $g_{\mu\nu} = g^{\mathrm{bg}}_{\mu\nu} + h_{\mu\nu}$.  This decomposition is not unique. The identification map that relates points of the background to points of the \ac{nr} spacetime can be changed without changing the physics. Therefore, part of $h_{\mu\nu}$ may be pure gauge, meaning that it describes the background itself in slightly shifted coordinates rather than a physical change of the geometry~\cite{Stewart:1974uz}. Separately, the choice of background mass and spin in this decomposition is not unique, so we also treat $M$ and $a$ as free parameters. The network then searches for the coordinate transformation, together with $M$ and $a$, that minimizes the difference between the pulled-back \ac{nr} metric and the background metric in the norm defined in Eq.~\eqref{eq:loss}. We are therefore effectively projecting the \ac{nr} solution onto the manifold of all Kerr metrics, rather than merely finding a coordinate transformation to a fixed metric. This projection is defined by the chosen norm and training domain. Adjusting the background parameters can absorb part of the physical perturbation into the background. We therefore compare the fitted mass and spin with the values obtained from the apparent horizon over the training interval in Sec.~\ref{sec:results.num.spin}. The residual describes the remaining departure from the fitted background in the selected gauge, together with numerical and fitting errors. Our aim is to quantify this residual and assess the recovery of the background parameters.

In full generality, identifying such a map is complicated by the fact that a given coordinate chart may cover only part of the spacetime manifold. However, in the context of \ac{nr} where numerical solutions are defined only on a compact domain, this complication can be sidestepped for this discussion. Additionally, the map is determined only up to isometries of the target spacetime.

\subsection{Neural Networks}

The problem posed above amounts to finding an unknown smooth map $\varphi_{\rm NN}$ between two coordinate systems. \ac{nn} are natural candidates for this task as they are capable of modeling complex and high-dimensional mappings. We briefly review the relevant background before describing our architecture and training procedure. 

\subsubsection{Feedforward Neural Networks}
Fully connected feedforward \acp{nn}, also known as multi-layer perceptrons, consist of three main components: an input layer, one or more hidden layers, and an output layer. In this fully connected architecture, every node (also called neuron) in a given layer receives input from the preceding layer, and calculates a weighted sum, adds a bias and applies an activation function.  Without the nonlinear activation, the composition of layers would
collapse to a single affine map; the activation is what lets the network represent nonlinear functions of its inputs. In feedforward \ac{nn}, the information propagates in one forward direction from the input layer through the hidden layers to the output layer. These \ac{nn} are trained via gradient-based optimization methods, such as stochastic gradient descent~\cite{10.1007/978-3-7908-2604-3_16}, which iteratively updates the weights and biases by computing the gradient of the loss function with respect to each parameter using backpropagation method~\cite{Rumelhart:1986gxv}. 

\subsubsection{Universal Approximators}
\ac{nn} are a class of universal approximators which suggests that a network with at least one hidden layer and a sufficient number of nodes under suitable assumptions on the activation function can approximate any continuous function on a compact domain to an arbitrary precision \cite{Cybenko:1989iql, Hornik:1989yye, FUNAHASHI1989183,HORNIK1990551,  2024arXiv240712895A}. It is precisely this capacity for flexible function approximation that motivates the use of \acp{nn} as a candidate framework for tackling the gauge problem. However, these
results do not guarantee that a chosen finite network
can be trained to achieve a desired accuracy, nor
do they determine the most effective architecture
or training procedure for a given problem.

Our application also requires derivatives of the coordinate map,
since the metric transformation depends on its Jacobian, and accurate
approximation of the coordinate values alone does not ensure accurate
derivatives. Extensions of the approximation results show that, for
sufficiently smooth activation functions, a network can approximate a
target function together with its derivatives on a compact
domain~\cite{HORNIK1990551}. As before, these results establish
existence only; they do not guarantee that training will find such a
network.
\subsubsection{Architecture \& Loss Function}

In this section, we provide a general overview of our \ac{nn} architecture and training procedure. A key feature of our approach is that the \ac{nn} is not trained to generalise across all \ac{nr} simulations; instead, each network is fitted to a single coordinate mapping. We never evaluate a trained network outside its training domain or apply it to data from other simulations.

\begin{figure*}
 \centering
  \includegraphics[width=1.0\textwidth]{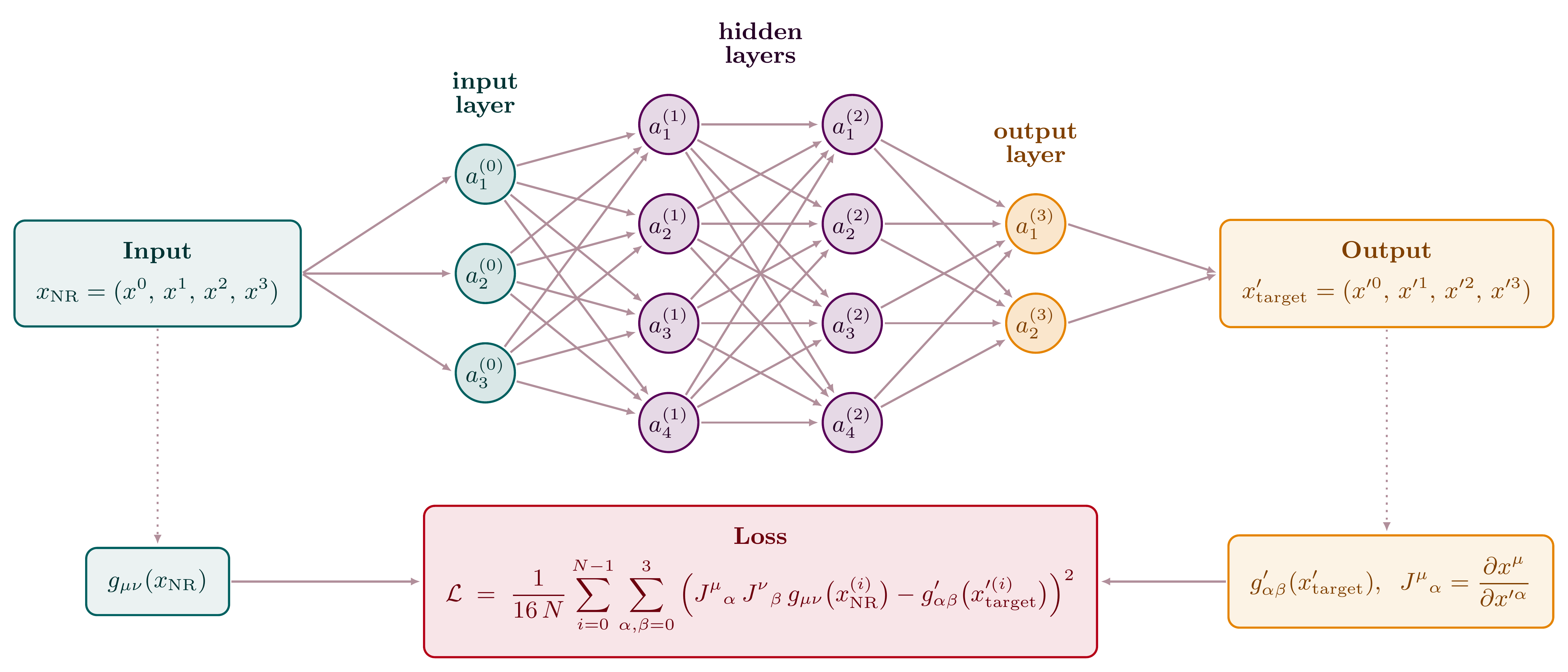}
 \caption{Schematic of the training workflow. The network takes the \ac{nr} coordinates $x_{\mathrm{NR}} = (x^0, x^1, x^2, x^3)$ as input and outputs the transformed coordinates $x'_{\mathrm{target}} = (x'^0, x'^1, x'^2, x'^3)$, parametrising the diffeomorphism between the two frames. The  metric in the \ac{nr} frame $g_{\mu\nu}(x_{\mathrm{NR}})$ and  the analytic target metric $g'_{\alpha\beta}(x'_{\mathrm{target}})$ are combined with the Jacobian $J^\mu{}_\alpha = \partial x^\mu / \partial x'^\alpha$, computed by automatic differentiation, to evaluate the pulled-back metric. The loss $\mathcal{L}$ penalises the squared difference between the pulled-back \ac{nr} metric and the target metric, summed over all tensor components and training points.}
  \label{fig:schematic}

 \end{figure*}

As shown schematically in Figure~\ref{fig:schematic}, the \ac{nn} takes as input the four source coordinates $x_{\mathrm{NR}}$ and returns the transformed coordinates $x'_{\mathrm{target}}$. To aid optimisation, the final stage of the network has a residual connection~\cite{Liu:2021azq,He:2015wrn}, so that the output is expressed as

\begin{equation}
    x'_{\mathrm{target}} = \varphi_{\mathrm{NN}}(x_{\mathrm{NR}})
    = x_{\mathrm{NR}} + f_{\theta}(x_{\mathrm{NR}}),
    \label{eq:residual}
\end{equation}
where $f_{\theta}$ is the output of the fully connected network with
trainable parameters $\theta$ (weights and biases). The network therefore learns only the displacement $f_\theta$ from the identity, while
$\varphi_{\mathrm{NN}}$ denotes the full coordinate map. This
parametrisation lets the map reduce exactly to the identity when
$f_\theta$ vanishes. We remark that by initializing the final layer to zero, we start training from the identity map and let the network learn the required coordinate displacement. Although this initialization provides a natural starting point, transformations involving large
departures from the identity may be more challenging to identify.

For applications to \ac{nr} data, the source metric
$g_{\mu\nu}(x_{\mathrm{NR}})$ is supplied by the
solvers described in Sec.~\ref{sec:setup}.
In the analytical tests, it is evaluated from the
known analytical spacetime metric. In the simulation data, each \ac{nr} coordinate has an associated numerical spacetime metric $g_{\mu\nu}(x_{\mathrm{NR}})$. The target coordinate system is specified by prescribing the functional form of the metric in the target frame, $g'_{\alpha\beta}(x'_{\mathrm{target}})$.
Using automatic
differentiation~\cite{Baydin:2015tfa},
we obtain the Jacobian matrix of the learned map,
$A^\alpha{}_\mu = \partial x'^\alpha/\partial x^\mu$,
and invert it to construct
\begin{equation}
    J^\mu{}_\alpha
    = \frac{\partial x^\mu}{\partial x'^\alpha}
    = (A^{-1})^\mu{}_\alpha,
    \label{eq:jacobian}
\end{equation}
which lets us evaluate the metric pulled back by the
inverse map through the standard tensor transformation
law (Eq.~\ref{eq:metric_transform}),
\begin{equation}
    g^{\mathrm{pullback}}_{\alpha\beta}
    \!\left(x'_{\mathrm{target}}\right)
    = J^\mu{}_\alpha\,J^\nu{}_\beta\,
      g_{\mu\nu}(x_{\mathrm{NR}}),
    \label{eq:metric_transform2}
\end{equation}
where $x'_{\mathrm{target}}=\varphi_{\rm NN}(x_{\mathrm{NR}})$. Note that throughout, indices $\mu,\nu$ refer to the \ac{nr} coordinates $x$ and $\alpha,\beta$ to the target coordinates $x'$; $A$ is the Jacobian of the
forward map $\varphi_{\rm NN}$ and $J=A^{-1}$ that of its inverse.

The network is trained by minimising a loss that
penalises the mismatch between the pulled-back metric
$g^{\mathrm{pullback}}_{\alpha\beta}$ and the prescribed
target metric $g'_{\alpha\beta}$,
\begin{equation}
    \mathcal{L} = \frac{1}{16\,N}\sum_{i=0}^{N-1}
    \sum_{\alpha,\beta=0}^{3}
    \left(
    g^{\mathrm{pullback}}_{\alpha\beta}
    \!\left(x'^{(i)}_{\mathrm{target}}\right)
    - g'_{\alpha\beta}
    \!\left(x'^{(i)}_{\mathrm{target}}\right)
    \right)^2,
    \label{eq:loss}
\end{equation}
where $N$ is the total number of points sampled from
the \ac{nr} data. Because the construction of $J^\mu{}_\alpha$ requires
inverting the Jacobian of the map, transformations
whose Jacobian determinant approaches zero drive
the loss to divergence, provided the source metric
remains nondegenerate and the target metric
components remain bounded. Minimising $\mathcal{L}$
therefore discourages such degeneracies at the
sampled points and favours locally invertible
coordinate transformations, although it does not
guarantee global invertibility.

The architecture is implemented in \texttt{PyTorch}. The number of hidden layers and the number of
neurons per layer vary between tests, as summarised in Table~\ref{tab:hyperparameters}.
We use the SiLU activation in the hidden
layers~\cite{elfwing2017sigmoidweightedlinearunitsneural}, defined by
\begin{equation}
    \operatorname{SiLU}(z)
    = \frac{z}{1+\exp(-z)}.
\end{equation}
SiLU is smooth ($C^\infty$), ensuring that the
network represents a smooth coordinate map whose
Jacobian varies smoothly with the inputs.
In contrast, a network with ReLU activations is
generally only piecewise differentiable.

Training is performed using the AdamW optimiser~\cite{Loshchilov:2017bsp} with an initial learning rate of $\eta_0 = 10^{-3} $ and an
exponential learning-rate scheduler (\texttt{ExponentialLR}). Mini-batch sizes and
training durations are chosen separately for each
test and are summarised in
Table~\ref{tab:hyperparameters}. Gradients of the
loss with respect to the network parameters are
computed by differentiating through the Jacobian
calculation, its inversion, and the metric
transformation. In the perturbed spinning
black-hole test, the target mass $M$ and spin
parameter $a$ are also optimised alongside the network parameters.
\begin{table*}
    \centering
    \caption{Neural-network architectures and training
    hyperparameters for the tests in
    Sec.~\ref{sec:results}. Batch sizes for the numerical
    tests are quoted per GPU.}
    \label{tab:hyperparameters}
    \begin{tabular}{lccccc}
        \hline\hline
        Test & Hidden layers & Nodes/layer
                & Activation & Batch size & Epochs \\
        \hline
        Schwarzschild: \ac{ef} $\to$ Schwarzschild
            & 3 & 128 & SiLU & 512  & 500 \\
        Kerr: BL $\to$ ingoing Kerr
            & 3 & 128 & SiLU & 512  & 500 \\
        Non-spinning puncture $\to$ EF
            & 4 & 128 & SiLU & 4096 & 2886 \\
        Spinning puncture $\to$ Cartesian KS
            & 4 & 512 & SiLU & 4096 & 366 \\
        \hline\hline
    \end{tabular}
\end{table*}

Training costs vary across the tests in
Table~\ref{tab:hyperparameters}. For the non-spinning
puncture model, training for $2886$ epochs on four
NVIDIA A100 GPUs required approximately $4$ hours,
corresponding to $16$ GPU-hours. This excludes
initialization, checkpoint writing, and data
preparation.

\section{Numerical Setup}
\label{sec:setup}
We train and evaluate our network on two \ac{nr} datasets of increasing difficulty. The first is a spherically symmetric \ac{bh}, whose symmetry reduces the simulation to $1{+}1$ dimensions. We then turn to the more demanding case of a perturbed spinning \ac{bh}, to test whether the \ac{nn} can identify a coordinate map between the background metric and the physical metric.

\subsection{GRHyps Solver}
\label{sec:setup.sch}
We evolve a single, non-spinning \ac{bh} in spherical symmetry using the Z4c formulation \cite{Bernuzzi:2009ex} with moving puncture gauge (Sec~\ref{sec:bg.punc}). 
The initial spatial geometry is an isotropic
Schwarzschild puncture slice,
\begin{equation}
    \dd l^2 = \psi^4\left(\dd r^2+r^2\dd \Omega^2\right),
    \qquad
    \psi=1+\frac{M}{2r},
\end{equation}
with initial lapse $\alpha=\psi^{-2}$ and vanishing
shift. We set the ADM mass to $M=1$.
Spatial derivatives are computed with a fourth-order finite differencing scheme, and time integration uses a fourth-order Runge--Kutta (RK) method. The radial domain extends to $r_{\max}=170M$, with uniform spacing $\Delta r=0.01M$ and a \ac{cfl} factor of $0.5$. Kreiss--Oliger dissipation is applied with coefficient $\epsilon_{\mathrm{KO}}=0.007$.

Although the underlying vacuum spacetime is Schwarzschild and therefore static, its representation in moving-puncture coordinates is time dependent. Starting from the initial isotropic puncture slice, the lapse
determines how the slices advance in time, while the shift moves the spatial coordinates within each slice. As described in
Sec.~\ref{sec:bg.punc}, the region sampled by the numerical grid approaches a portion of the $1+\log$ trumpet slice~\cite{Brown:2007tb,Hannam:2008sg}. During this evolution, the metric components at fixed numerical radius vary with time, and the
apparent horizon moves outward in coordinate radius while its areal radius remains approximately $2M$. These changes reflect gauge dynamics in a spacetime without physical gravitational radiation. This provides a nontrivial test of the learned map: the network must account for the evolving slicing
and radial coordinates to recover a stationary
target metric throughout the training interval.

In spherical symmetry, the 3+1 spacetime metric (Eq. \ref{eq: ADMmetric}) reduces to
\begin{equation}
    \dd s^2 = -(\alpha^2-\beta_{r}\beta^{r})\,\dd t^2 + 2\,\beta_{r}\,\dd r\,\dd t + \gamma_{rr}\,\dd r^{2} + \gamma_{T}\,r^2\,\dd\Omega^{2},
\end{equation}
where $\beta_{r}=\beta^{r}\gamma_{rr}$ and
$\dd\Omega^2=\dd\theta^2+\sin^2\theta\,\dd\phi^2$. We extract the lapse $\alpha$, radial shift $\beta^r$,
and spatial-metric functions $\gamma_{rr}$ and
$\gamma_T$ on the numerical $(t,r)$ grid.
These functions are independent of angle.
The four-dimensional metric is reconstructed from
them using the spherical coordinate basis, including
the angular dependence
$g_{\phi\phi}=r^2\gamma_T\sin^2\theta$.

To train the full four-dimensional coordinate map,
we augment the numerical data with angular
coordinates $(\theta,\phi)$.
We use $101$ time slices spanning
$0\leq t\leq50M$, separated by $0.5M$.
On each slice, we retain every second stored radial
point outside the apparent horizon, subject to
$r\leq10M$, giving a radial sampling spacing of
$0.02M$. We exclude points inside the apparent horizon from
the training data. Because the horizon's coordinate
location changes during the gauge evolution, the
inner radial cutoff must also vary with time:
we retain only points with $r\geq r_{\mathrm{AH}}(t)$.
We determine $r_{\mathrm{AH}}(t)$ by interpolating
the horizon-finder output to the sampled times.
Its value increases from approximately $0.50M$
to $1.18M$ over the training interval.

For each retained $(t,r)$ point, we independently
draw $N_\theta=10$ polar angles using
$\theta=\arccos u$, with
$u\sim\mathcal{U}(-1,1)$, and $N_\phi=20$
azimuthal angles from $\mathcal{U}[0,2\pi)$.
Their tensor product gives $200$ angular pairs per
spacetime point.
The angular samples are generated once during data
preparation and held fixed during training.
The resulting training set contains
$\sim 9 \times 10^{6}$ four-dimensional points.

\subsection{AthenaK Solver}
\label{sec:setup.kerr}
Our second numerical dataset describes the evolution
of a perturbed, rotating black hole. We generate this dataset with \texttt{AthenaK}~\cite{Zhu:2024utz}, which implements the Z4c evolution. Spatial derivatives are evaluated with sixth-order finite differences, using four ghost cells, and the evolution is advanced with a fourth-order
Runge--Kutta scheme.

We obtain the initial data with
\texttt{TwoPunctures}~\cite{Ansorg:2004ds}, which solves
the Hamiltonian constraint spectrally for
conformally flat puncture data.
To describe an approximately isolated rotating
black hole within a two-puncture configuration,
we assign almost all the mass and all the spin
to one puncture.
The primary has target mass $M_1=1M$ and
dimensionless spin $\chi=J/M_1^2=0.5$.
A nonspinning secondary with mass $10^{-12}M$
is placed at a coordinate separation of $10^{-5}M$.
Both the separation and the secondary's mass scale
are too small for it to appear as a distinct
resolved object in the evolution.
This prescription follows the spinning-black-hole
tests in \texttt{BAM}~\cite{Hilditch:2012fp}, \texttt{GR-Athena++}~\cite{Daszuta:2021ecf} and \texttt{AthenaK}~\cite{Zhu:2024utz}. The computational domain is a cube with $x,y,z\in[-1024M,1024M]$.
Static refinement concentrates the spatial
resolution around the black hole, where the
smallest grid spacing is approximately $0.0833M$.
We choose a \ac{cfl} factor of
$0.25$, corresponding to a timestep of approximately
$0.0208M$, and apply Kreiss--Oliger dissipation
with coefficient $\epsilon_{\mathrm{KO}}=0.2$. 

The initial data's departure from Kerr is not solely a consequence of introducing the secondary puncture. The Bowen--York construction used by \texttt{TwoPunctures} assumes a conformally flat initial spatial metric, $\gamma_{ij}=\psi^4\delta_{ij}$, with spin encoded in the extrinsic curvature. For a rotating black hole, this conformal-flatness assumption is incompatible with a stationary Kerr slice under the conditions discussed in Ref.~\cite{Garat:2000pn,ValienteKroon:2003ux}. The resulting data therefore describes a perturbed black hole that emits gravitational radiation as
it relaxes toward Kerr~\cite{Gleiser:1997ng}. After $t \approx 40\,M$ the apparent horizon settles to a Kerr black hole with mass $M_{\rm H} = 1.0566$ and dimensionless spin $\chi = J_{\rm H}/M_{\rm H}^2 = 0.4596$ ($a = J_{\rm H}/M_{\rm H} = 0.4856$). Here $M_{\rm H}$ and $J_{\rm H}$ are the quasi-local isolated-horizon mass and angular momentum~\cite{Dreyer:2002mx,Ashtekar:2004cn}, which remain nearly constant ($\lesssim 10^{-4}$) for the rest of the evolution.

The network must therefore learn a coordinate map
that accounts for the evolving numerical gauge
while identifying an approximate Kerr background.
Physical deviations from Kerr remain as a residual
metric mismatch that cannot be removed by a
coordinate transformation. For training, we recover the physical spatial
metric $\gamma_{ij}$ from the conformal variables
stored by the evolution code.
Together with the lapse and shift, these fields
determine the four-dimensional metric through
Eq.~\eqref{eq: ADMmetric}.
We use $198$ snapshots between $t=0.625M$ and
$t=62.1875M$, with a spacing of $0.3125M$. Static mesh refinement produces a nonuniform radial distribution of grid points, with denser resolution near the puncture. We therefore use a hybrid sampling procedure to select training data points. We sample the spatial grid according to the
coordinate radius $r=\sqrt{x^2+y^2+z^2}$.
At each time, all active grid points satisfying
$r_{\min}(t)\leq r<3M$ are included. For $3M\leq r\leq100M$, we instead divide the radial range into $100$ logarithmically spaced
shells and select $1000$ points at random from
each shell, or all its points if fewer are available.
Keeping the inner grid intact preserves its strong-field resolution, while sampling the
outer shells prevents their larger populations from dominating the training set. The inner radial sampling boundary follows the changing
coordinate size of the apparent horizon. Specifically, we interpolate the smallest coordinate radius reported by the horizon finder to obtain $r_{\min}(t)$ at each snapshot. This defines a spherical cutoff; however the horizon is not spherical, some points inside the horizon may remain in the selected data. We randomly sample points at each time slice and the complete dataset contains $\sim5.7 \times 10^7$ spacetime points.
 
\subsection{Teukolsky Solver}
\label{sec:setup.teuk}
At sufficiently late times in the evolution described in Sec.~\ref{sec:setup.kerr}, we expect the numerical spacetime to be well approximated by a linear perturbation of Kerr. Its radiative content can then be evolved using the Teukolsky equation~\cite{Teukolsky:1973ha}. We use the learned map to construct Cauchy initial
data from the \ac{nr} solution on several slices of
constant \ac{ks} time, evolve these data with a linear
solver, and compare the resulting waveforms with those
extracted directly by \texttt{AthenaK}. This comparison
tests whether the learned background identification
preserves the radiative content of the numerical
spacetime.

\paragraph{Teukolsky equation in \ac{ks} coordinates.}

The radiative degrees of freedom are encoded in the Weyl tensor $C_{\mu\nu\rho\sigma}$, which coincides with the Riemann tensor in vacuum.
Given a null tetrad $(\ell^\mu, n^\mu, m^\mu, \bar m^\mu)$, we use
\begin{equation}
\psi_4 = -C_{\mu\nu\rho\sigma}\, n^\mu \bar m^\nu n^\rho \bar m^\sigma.
\label{eq:psi4-def}
\end{equation} 

The Kerr spacetime is Petrov type D, and its background
$\psi_4$ vanishes in a tetrad aligned with the principal
null directions. Because the background values vanish,
the first-order perturbation of $\psi_4$ is invariant
under infinitesimal coordinate and tetrad transformations~\cite{Teukolsky:1973ha}. The remaining ambiguity in the identification discussed in Sec.~\ref{sec:bg.gauge} therefore does not affect $\psi_4$ at first order, provided that the learned map has removed the large gauge difference between the NR coordinates and the background, and that the remaining identification and tetrad differences are of perturbative order. The fitted mass and spin enter differently: they specify the background on which $\psi_4$ is evolved and hence the quasinormal frequencies of the Teukolsky evolution. 

We use the horizon-regular formulation of the Teukolsky equation in ingoing \ac{ks} coordinates~\cite{Campanelli:2000nc}. $\psi_4$ is rescaled and decomposed into azimuthal modes,
\begin{equation}
    \psi
    = (r-ia\cos\theta)^4\psi_4
    = \sum_{m\in\mathbb{Z}}
      \psi_m(\tilde{t},r,\theta)
      e^{im\tilde{\varphi}}.
      \label{eq:psi4-scale}
\end{equation}

Each mode satisfies the Teukolsky equation in ingoing \ac{ks} coordinates $(\tilde{t},r,\theta,\phi)$ \cite{Campanelli:2000nc},
\begin{equation}
\begin{aligned}
    &(\Sigma+2Mr)
      \frac{\partial^2\psi_m}{\partial\tilde{t}^{\,2}}
    - \Delta
      \frac{\partial^2\psi_m}{\partial r^2}
    - (2iam+6r-6M)
      \frac{\partial\psi_m}{\partial r}
    \\[2pt]
    &\quad
    - \frac{1}{\sin\theta}
      \frac{\partial}{\partial\theta}
      \left(
          \sin\theta
          \frac{\partial\psi_m}{\partial\theta}
      \right)
    - 4Mr
      \frac{\partial^2\psi_m}
           {\partial\tilde{t}\,\partial r}
    \\[2pt]
    &\quad
    - (4r+4ia\cos\theta+6M)
      \frac{\partial\psi_m}{\partial\tilde{t}}
    \\[2pt]
    &\quad
    + \left(
        4\cot^2\theta - 2
        + m^2\csc^2\theta
        - 4m\cot\theta\,\csc\theta
      \right)\psi_m
    = 0.
\end{aligned}
\label{eq:teuk_ks}
\end{equation}
where, 
\begin{equation}
    \Sigma = r^2 + a^2\cos^2\theta,
    \qquad
    \Delta = r^2 - 2Mr + a^2.
\end{equation} 
The scalar $\psi_4$ in Eq.~\eqref{eq:psi4-def} is evaluated in the Kinnersley tetrad boosted by $\Delta$ ($\ell = \Delta\,\ell_{\rm K}$, $n = \Delta^{-1} n_{\rm K}$, $m = m_{\rm K}$, so that $\psi_4 = \Delta^{-2}\psi_4^{\rm K}$), whose components in the ingoing \ac{ks} basis $(\tilde t, r, \theta, \phi)$ are
\begin{equation}
\begin{aligned}
\ell^\mu &= \left( \Delta+4Mr,\, \Delta,\, 0,\, 2a \right),
\\
n^\mu &= \frac{1}{2\Sigma} \left(1,\,-1,\,0,\,0\right),
\\
m^\mu &= \frac{1}{\sqrt{2}(r+ia\cos\theta)}
\left( ia\sin\theta,\, 0,\, 1,\, i\csc\theta \right).
\end{aligned}
\label{eq:tetrad}
\end{equation}
With metric signature $(-,+,+,+)$ it satisfies $\ell^\mu n_\mu = -1$ and $m^\mu \bar m_\mu = 1$, and all other inner products vanish. Unlike the Kinnersley tetrad, the boosted tetrad is regular across the future horizon, so $\psi$ remains finite on the inner boundary of the evolution domain. 

In our implementation, the \ac{nn} outputs Cartesian \ac{ks} coordinates $( t', X', Y', Z')$ that are related to the ingoing \ac{ks} coordinates by
\begin{equation}
X' + iY' = (r + ia)\sin\theta\, e^{i\phi},
\qquad
Z' = r\cos\theta.
\label{eq:ks-cartesian}
\end{equation}
We therefore transform the analytic tetrad components
to this Cartesian basis before applying the learned
coordinate map.

\paragraph{Computing $\psi_4$ from the \ac{nr} data.}
The Teukolsky solver needs $\psi$ and its time derivative on its \ac{ks} grid, but the numerical metric data is given in \ac{nr} coordinates. Computing curvature after transforming the \ac{nr} metric to \ac{ks} coordinates would require higher derivatives of the network which can be noisy. We instead compute the curvature directly in the \ac{nr} coordinates. For a specified tetrad, $\psi_4$ is a scalar. The learned map is then needed for two things: to find the \ac{nr} point corresponding to each \ac{ks} grid point, and to express the tetrad (Eq.~\ref{eq:tetrad}) in \ac{nr} coordinates.

The first of these requires inverting the learned map $\varphi_{\rm NN}$. It maps \ac{nr} coordinates $x^\alpha$ to \ac{ks} coordinates $x'^{\alpha} = \varphi_{\rm NN}^{\alpha}(x)$. Applied to the \ac{nr} grid, it would give a scattered set of \ac{ks} points that do not lie on the Teukolsky solver's grid, and the data would have to be interpolated between
these scattered points. We instead start from the solver's \ac{ks} grid and, for each grid point $x'_\ast$, find the \ac{nr} point that maps onto it. All interpolation then takes place on the structured \ac{nr} mesh. For each $x'_\ast$ we solve $\varphi_{\rm NN}(x) = x'_\ast$ for $x$ with
Newton's method. The forward-map Jacobian,
\begin{equation}
    A^\alpha{}_\mu
    =\frac{\partial\varphi_{\rm NN}^\alpha}
           {\partial x^\mu},
\end{equation}
is obtained by automatic differentiation. A point is accepted only if the residual $\lVert \varphi_{\rm NN}(x) - x'_\ast\rVert$ falls below a fixed tolerance and the recovered point $x$ lies inside the training domain described in
Sec.~\ref{sec:setup.kerr}. 

On each stored \ac{nr} snapshot we recover the spatial metric $\gamma_{ij}$ and extrinsic curvature $K_{ij}$ from the Z4c variables and compute the projections of the Riemann tensor onto the slice and its normal, following \cite{Bruegmann:2006ulg}. Similar to \texttt{AthenaK}, spatial derivatives use sixth-order finite differencing. We then interpolate these projections, together with the lapse and shift, to each \ac{nr} point $x$ found above. We use sixth-order Lagrange polynomials in space and time, using the same grid points for all five time slices (see below) so that the time derivative is not affected by change of stencil.

We evaluate the analytic tetrad at $x'_\ast$ and
transform its vectors to \ac{nr} coordinates using
the inverse forward-map Jacobian:
\begin{equation}
    v^\mu
    =\left(A^{-1}\right)^\mu{}_\nu v'^\nu,
    \qquad
    v\in\{n,\bar m\}.
\end{equation}
Here $v'^\nu$ denotes the Cartesian \ac{ks}
components and $v^\mu$ the \ac{nr} components.
Contracting the interpolated curvature with these
vectors gives $\psi_4$, which is then rescaled
according to Eq.~\eqref{eq:psi4-scale}.

\paragraph{Numerical evolution.}
The solver is initialized with $\psi_m$ and $\partial_{t'}\psi_m$ on a slice $t' = t_0$. The time derivative is computed using a fourth-order centered difference of $\psi_m$ using five slices $t_0 + kh$, $k \in \{-2,\dots,2\}$, with $h = 0.01\,M$. Each azimuthal mode is obtained from $N_\phi$ equally spaced angles
$\phi_k = 2\pi k/N_\phi$,
\begin{equation}
    \psi_m = \frac{1}{N_\phi}\sum_{k=0}^{N_\phi-1}
      \psi(\phi_k)\, e^{-im\phi_k}.
    \label{eq:m-projection}
\end{equation}
In the evolutions presented here, we retain the axisymmetric mode $m=0$.

Finally, we evolve $\psi_m$ with
\texttt{PyTeukolsky}~\cite{pyteukolsky}, a finite-difference, time-domain solver. We run each case at two resolutions,
$(N_r, N_\mu) = (960, 64)$ and $(480, 32)$, with
$N_\phi=128$ and $64$, respectively, for the initial
azimuthal projection. The radial grid is uniform in
$x = \ln(r/M)$, with $N_r$ cells between $r_{\min}$ and
$r_{\max} = 400\,M$, so that
\begin{equation}
    r_i = r_{\min}\, e^{(i+1/2)\Delta x},
    \qquad
    \Delta x = \frac{1}{N_r}\ln\frac{r_{\max}}{r_{\min}},
\end{equation}
for $i = 0, \dots, N_r - 1$. This concentrates radial resolution near the puncture. The inner boundary lies just inside the outer horizon,
$r_{\min} = 0.9426\, r_+$, with $r_+ = M + \sqrt{M^2 - a^2}$. The polar grid
is uniform in $\mu = \cos\theta$ and staggered, so that no grid point lies
on a pole. Radial derivatives are fourth-order and angular derivatives are second-order accurate.

Although the nominal radial grid extends inside
the horizon, points with $r\leq r_+$ are excised
from the active evolution. Near the horizon,
outward one-sided radial stencils avoid accessing
the excised points.  At the
outer boundary we apply a Sommerfeld condition, which is far away that
reflections cannot reach the detectors during the run. We add Kreiss--Oliger dissipation with $\epsilon_{\rm KO}=0.025$, disabling its radial contribution wherever the dissipation stencil would enter the excised region. We integrate in time with fourth-order Runge--Kutta at a CFL factor of $0.4$ for $200\,M$.

We extract the waveform at $r = 60\,M$ and $80\,M$ by projecting $\psi_4$ onto the spin-weighted spherical harmonics ${}_{-2}Y_{\ell m}$. Although \texttt{AthenaK} uses the same
sign convention and harmonics, it evaluates $\psi_4$ in a
different null tetrad. Its tetrad is constructed from the
future-directed unit normal $\hat n$ to the \ac{nr} slice
and a spatial triad $(\hat r,\hat\theta,\hat\varphi)$,
orthonormalized using the Gram--Schmidt
procedure~\cite{Bruegmann:2006ulg}:
\begin{equation}
    n_{\rm A}=\frac{\hat n-\hat r}{\sqrt{2}}, \qquad
    l_{\rm A}=\frac{\hat n+\hat r}{\sqrt{2}}, \qquad
    \bar m_{\rm A}
    =\frac{\hat\theta-i\hat\varphi}{\sqrt{2}}.
\end{equation}
This construction follows the same prescription as the numerical tetrad used in the Lazarus approach~\cite{Campanelli:2005ia}.

Our $\psi_4$ is evaluated in the horizon-regular
boosted Kinnersley tetrad of
Eq.~\eqref{eq:tetrad}. To compare with the \texttt{AthenaK} waveform,  we first undo this boost:
\begin{equation}
    \psi_4^{\rm K}=\Delta^2\psi_4.
\end{equation}
We then apply an approximate conversion from the
unboosted Kinnersley tetrad to the numerical tetrad,
using the Kerr-background transformation in
Eq.~(2.15) of
Ref.~\cite{Campanelli:2005ia}.
That transformation is derived for a numerical tetrad on constant \ac{bl}-time slices of Kerr. It includes null rotations, which mix the Weyl
scalars, and a spin-boost, which changes the normalization and phase of $\psi_4$. At large radius, the rotation parameter $A\simeq a\sin\theta/r$ becomes small and thus the
mixing contributions can be ignored in the radiation
zone. In other words, in regions far from the strong-field, we can assume that the numerical tetrad is close to Kinnersley tetrad's null direction \cite{Nakano:2015pta}. We therefore neglect these contributions at our extraction radii and retain the spin-boost conversion 
\begin{equation}
    \psi_4^{\rm NR}
    \approx F_A^2F_B^2\,\psi_4^{\rm K}
    =\frac{2\zeta^2}{\Delta}\,\psi_4^{\rm K},
    \label{eq:tetrad-factor}
\end{equation}
where
\begin{align*}
    F_A &= \sqrt{\frac{2\Sigma}{\Delta}},
    & F_B &= \frac{\sqrt{\Sigma}}{\bar\zeta}, \\
    \zeta &= r-ia\mu,
    & \mu &= \cos\theta.
\end{align*}

Combining the boost correction with Eq.~\eqref{eq:tetrad-factor}, we get
\begin{equation}
    \psi_4^{\rm NR}\approx 2\Delta\zeta^2\,\psi_4
\end{equation}
for the waveform comparison. This angle-dependent conversion is applied before the spherical-harmonic projection. We highlight that the conversion remains approximate at finite radius: it neglects the mixing with the other Weyl scalars, and it assumes that \texttt{AthenaK}'s
slices near the extraction sphere are close to slices of constant \ac{bl} time, as in the derivation of Ref.~\cite{Campanelli:2005ia}.

\section{Results}
\label{sec:results}
In this section we present a series of tests that probe our methodology. We begin with analytical cases, in which the closed form of the coordinate transformation is known a priori. These tests serve as controlled benchmarks; the ground-truth map is available for us to directly assess how faithfully the network recovers the underlying transformation and quantify its accuracy. We then proceed to more challenging settings where no closed form map exists.

\subsection{Analytical Cases}
\label{sec:results.analyt}
\subsubsection{Schwarzschild Test}
\label{sec:results.analyt.sch}

We start by training our \acs{nn} to discover a map from ingoing \ac{ef} coordinates $(v,r,\theta,\phi)$, where $v$ is the advanced null coordinate, to Schwarzschild coordinates $(t',r',\theta',\phi')$ in spherical polar form. The explicit forms of the metric in both coordinate systems are given in Appendix~\ref{app:metrics}. The analytical transformation is,
\begin{align}
    t' &= v-\int \frac{dr}{1-2M/r}\,=\,v - r - 2M\ln\left|\frac{r}{2M}-1\right|, \nonumber \\
    r' &= r, \qquad \theta' = \theta, \qquad \phi' = \phi.
\label{eq:schd_analytic}
\end{align}
Only the time coordinate is modified while the radial and angular coordinates are identity transformations. We assume $M=1$. Since the target Schwarzschild metric possesses a coordinate singularity at the horizon $r=2$M, we restrict sampling our data points well outside the horizon to avoid this region. Specifically, we sample the input coordinates on a uniform grid, with $v \in [3,10]$, $r \in [4,12]$, $\theta \in (0,\pi)$, and $\phi \in [0,2\pi)$, taking a total of $ 3.9\times10^{5}$ points for training. 

\begin{figure}
    \centering
    \includegraphics[width=0.45\textwidth]{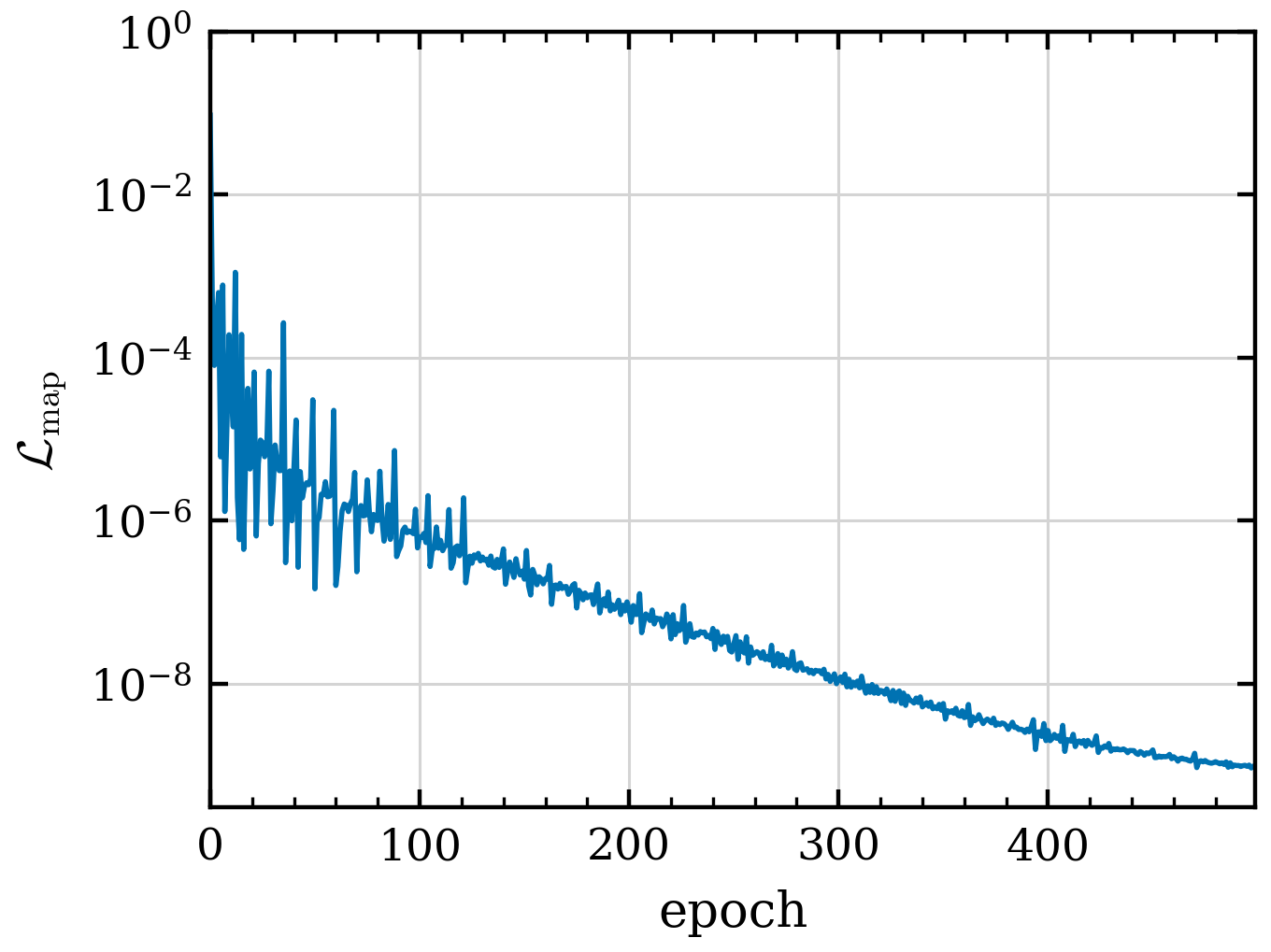}
     \caption{Training loss $\mathcal{L}_{\rm map}$ \eqref{eq:loss} for the map from ingoing \ac{ef} to Schwarzschild coordinates, shown as a function of epoch on a logarithmic scale. The loss measures the mean-squared difference between the target metric and the source metric transformed into the target coordinates. Following an initial noisy transient, it decreases to $\mathcal{O}(10^{-9})$ over $500$ epochs, indicating that it is able to learn the transformation.}
    \label{fig:ef_sc_loss}
\end{figure}

The network architecture and training hyperparameters are summarized in Table~\ref{tab:hyperparameters}. The model is trained for $500$ epochs by minimizing the mean-squared error (MSE) between the target Schwarzschild metric and the metric obtained by pulling the ingoing \ac{ef} metric back into the target coordinates through the learned transformation \eqref{eq:loss}. The evolution of this loss is shown in Fig.~\ref{fig:ef_sc_loss}. After a noisy transient over the first $\sim\!120$ epochs, the loss settles into a steady decline, falling from $\mathcal{O}(10^{-1})$ at initialization to $\mathcal{O}(10^{-9})$ by the end of training, a reduction of roughly eight orders of magnitude.

\begin{figure}
    \centering
    \includegraphics[width=0.45\textwidth]{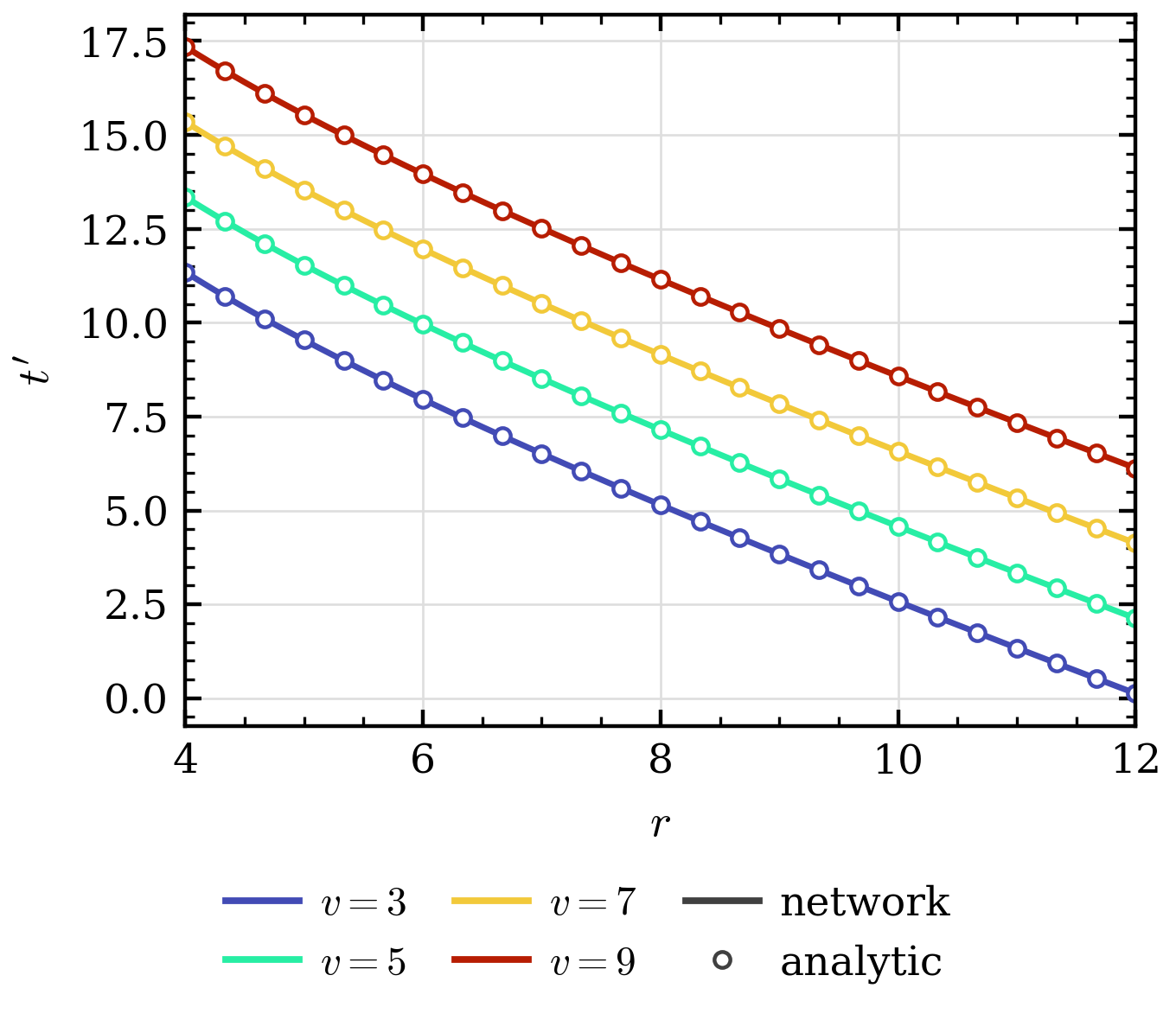}
        \caption{Learned Schwarzschild time coordinate $t'$ as a function of input $r$ for four fixed values of the advanced time $v\in\{3,5,7,9\}$, with $\theta=\pi/2$ and $\phi=\pi$. Solid lines show the network prediction and open circles represent the analytic transformation of Eq.~\eqref{eq:schd_analytic}. The analytical values are shifted by a single  constant $C_t\simeq12.35$, reflecting the time translation freedom of the target metric. After this shift, the predictions and analytical values agree within the resolution of the plot.}

    \label{fig:ef_sc_coord}
\end{figure}

The Schwarzschild metric is invariant under time translations and rotations, so metric matching determines transformation only up to these isometries. We find that for the learned solution, the difference between $t'$ and Eq.~\eqref{eq:schd_analytic} is approximately constant. Over the training grid this constant offset was measured to be $C_t \sim 12.35 \pm 5.3\times 10^{-5}$. This offset corresponds to the integration constant in the analytical transformation and has no physical significance. In Fig.~\ref{fig:ef_sc_coord} we shift the analytic curves by this constant, after which the prediction overlaps the analytic result at every plotted point. Similarly, the learned $\phi'$ differs from $\phi$ by a constant rotation $C_{\phi} \sim 3\times10^{-3} \pm 5.9\times 10^{-7}$, which corresponds to a rotation about the polar axis. 

To assess the accuracy independently of these additive
offsets, we compare the forward Jacobian of the learned
map, $A=\partial x'/\partial x$, with that of the
analytical transformation. We define the point-wise
relative error as
\begin{equation}
    \epsilon_A=
    \frac{\lVert A_{\rm NN}-A_{\rm exact}\rVert_F}
         {\lVert A_{\rm exact}\rVert_F},
    \label{eq:analytic_jacobian_error}
\end{equation}
where
$\lVert A\rVert_F=
(\sum_{\alpha\mu}A_{\alpha\mu}^2)^{1/2}$
is the Frobenius norm. Over the training grid, $\epsilon_A$ has a root-mean-square value of $4\times10^{-5}$ and a maximum of $7\times10^{-4}$. The maximum occurs at a corner of the sampled domain near a coordinate pole.

\subsubsection{Kerr Test}
\label{sec:results.analyt.kerr}
Next, we consider a spinning \ac{bh}. We train our model to map the \ac{bl} coordinates $(t,r,\theta,\phi)$ of the Kerr spacetime to the corresponding ingoing Kerr (IK) coordinates $(t',r',\theta',\phi')$~\cite{Teukolsky:2014vca,Visser:2007fj}. We take a black hole of mass $M=1$ and spin parameter $a/M=0.5$. IK and BL coordinates are related by,
\begin{gather}
    t' = t+ \int dr\, \frac{a^{2}+r^{2}}{r^2-2Mr+a^2}, \nonumber \\
    \phi' = \phi +\int dr\, \frac{a}{r^2-2Mr+a^2},
\label{eq:kerr_analytic}
\end{gather}
while the radial and polar coordinates are unchanged, $r'=r$ and $\theta'=\theta$. 
As in the Schwarzschild case, the \ac{bl} coordinates are singular at the horizons; with $a=0.5$ the outer horizon lies at $r_+\approx1.87$, so we again sample well outside it, on a uniform grid with $t\in[3,10]$, $r\in[4,12]$, $\theta\in(0,\pi)$, and $\phi\in[0,2\pi)$, taking a total of $3.9\times10^{5}$ points. 

\begin{figure}
    \centering
    \includegraphics[width=0.45\textwidth]{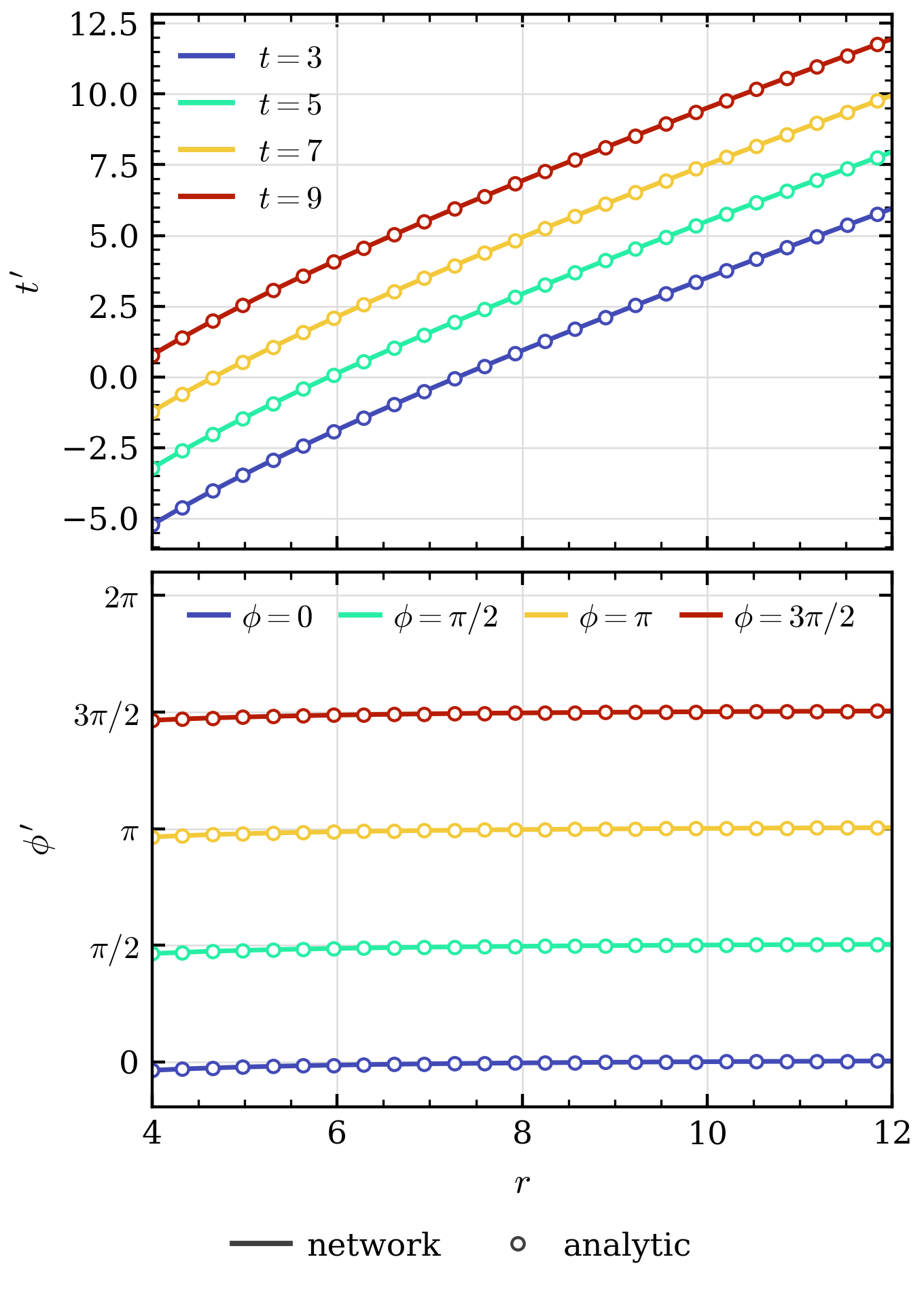}
    \caption{Learned ingoing Kerr coordinates as functions of $r$ for the map from Boyer--Lindquist coordinates ($M=1$, $a=0.5$). Top: $t'$ for four fixed values $t\in\{3,5,7,9\}$, with $\theta=\pi/2$ and $\phi=\pi$. Bottom: $\phi'$ for four fixed values $\phi\in\{0,\pi/2,\pi,3\pi/2\}$, with $t=5$ and $\theta=\pi/2$. Solid lines show the network prediction and open circles the analytic transformation of Eq.~\eqref{eq:kerr_analytic}. The analytical coordinates are shifted by
    $C_t\simeq-13.65$ and $C_\phi\simeq0.06$,
    using the same constants for every slice.
    After these shifts, the predictions and
    analytical values agree within the resolution
    of the plot.}
    \label{fig:bl_ef_coord}
\end{figure}

The network architecture and training settings are identical to the Schwarzschild case (Sec.~\ref{sec:results.analyt.sch}). We observe that the loss function drops to $\mathcal{O}(10^{-9})$ indicating that the network is learning a coordinate transformation. Figure~\ref{fig:bl_ef_coord} compares the two non-trivial learned coordinates as functions of $r$. The top panel shows $t'$ for four fixed values of input $t\in\{3,5,7,9\}$, at fixed $\theta=\pi/2$ and $\phi=\pi$, and the bottom panel displays $\phi'$ for four fixed values $\phi\in\{0,\pi/2,\pi,3\pi/2\}$, at fixed $t=5$ and $\theta=\pi/2$. In both panels the network prediction (solid lines) overlaps the analytic transformation of Eq.~\eqref{eq:kerr_analytic} (circles) across the full sampled range, $r\in[4,12]$, with no visible deviation for any slice.

The Kerr metric is invariant under time translations and rotations about the spin axis, so the loss constrains $t'$ and $\phi'$ only up to additive constants Eq.~\eqref{eq:kerr_analytic}. We determine these constants as the mean offset between the learned and analytic coordinates over the training grid, finding $C_t = -13.6524$ and $C_\phi = 0.0596$, with standard deviations of $4\times10^{-5}$ and $3\times10^{-6}$, respectively. We apply these same two constants to all analytical
curves in Fig.~\ref{fig:bl_ef_coord}. The radial and polar coordinates show mean deviations from the analytical map of $\mathcal{O}(10^{-6})$ or smaller.

Using the Jacobian error defined in
Eq.~\eqref{eq:analytic_jacobian_error}, we find a root-mean-square error of $5\times10^{-5}$ and a maximum of $2\times10^{-3}$ over the training grid. The maximum again occurs at a corner of the sampled domain near a coordinate pole. These errors are of comparable magnitude to those in the Schwarzschild case, demonstrating accurate recovery also for the spinning spacetime.

\subsection{Numerical Cases}
\label{sec:results.num}

\subsubsection{1+1D Puncture}
\label{sec:results.num.1d}

\begin{figure}
    \centering
    \includegraphics[width=0.45\textwidth]
        {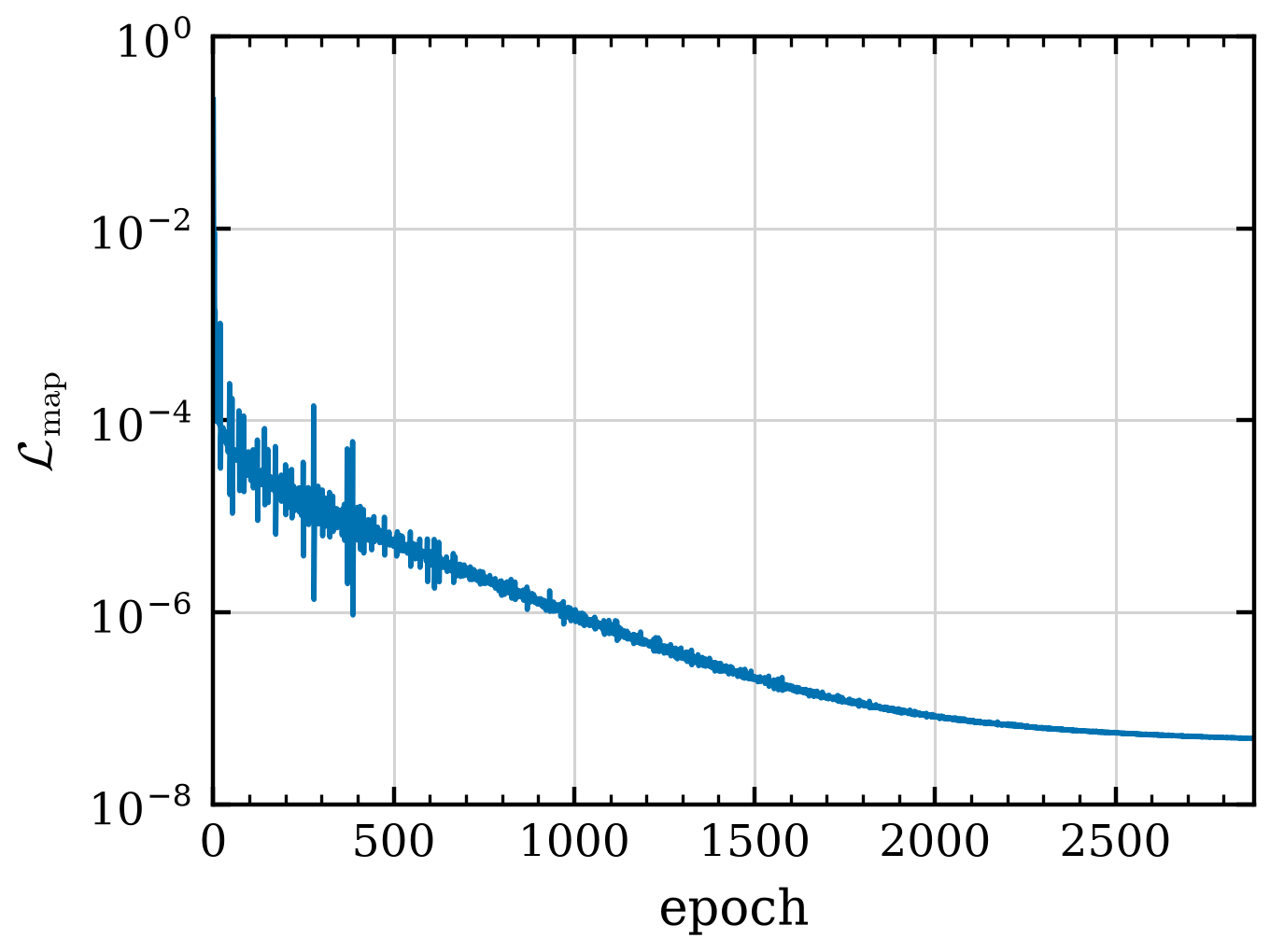}
    \caption{
    Mapping loss for the transformation from
    numerical $1+1$D puncture coordinates to ingoing
    \ac{ef} coordinates.
    Following an initial noisy transient, the loss decreases to approximately
    $4.8\times10^{-8}$ after $2886$ epochs.
    }
    \label{fig:1D_to_ef_loss}
\end{figure}

We apply the method to the non-spinning puncture
simulation described in Sec.~\ref{sec:setup.sch}.
The training domain covers
$0\leq t\leq50$ and
$r_{\rm AH}(t)\leq r\leq10$, where $(t,r)$ denote the numerical time and radial coordinates. The network maps the numerical coordinates
$(t,r,\theta,\phi)$ to ingoing
\ac{ef} coordinates
$(t',r',\theta',\phi')$, where $t'$ is the
advanced time and $r'$ is the areal radius.
The target metric is given in
Appendix~\ref{app:metrics}.
We fix its mass to $M=1.000016$, the irreducible mass determined from the apparent-horizon area at the final simulation time. We train it for
${\sim} 3000$ epochs, during which the loss function \eqref{eq:loss} decreases steadily from $\mathcal{O}(10^{-1})$ to $\mathcal{O}(10^{-8})$, as shown in Fig.~\ref{fig:1D_to_ef_loss}. We stop the training once the loss curve flattens and further epochs no longer reduce it appreciably. We remark that this coordinate transformation is non-trivial as the \ac{nr} simulation exhibits complex gauge  dynamics (see Sec.~\ref{sec:setup.sch}), whereas the metric components in the target \ac{ef} coordinates are time-independent. 

\begin{figure*}
    \centering
    \includegraphics[width=\textwidth]{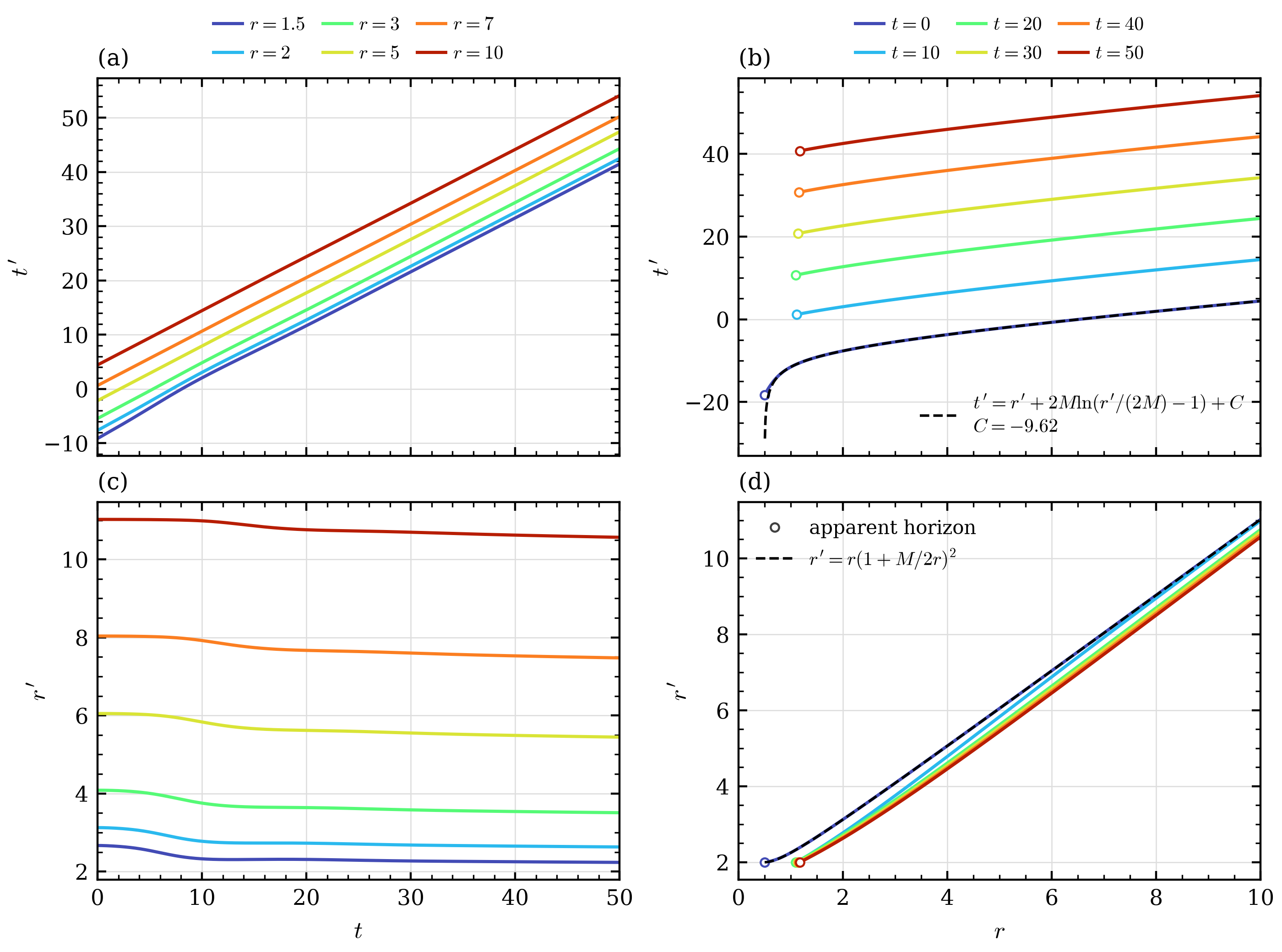}
    \caption{
    Learned time and radial maps at fixed input angles $(\theta,\phi)=(\pi/2,\pi)$. Panels~(a) and~(c) show $t'$ and $r'$ as functions of numerical time at the fixed
    input radii indicated in the left legend. Panels~(b) and~(d) show their dependence on input radius at the times indicated in the
    right legend. Open circles mark the location of the mapped apparent horizon. The dashed black curves show the initial-slice relations in Eqs.~\eqref{eq:puncture_initial_radius} and \eqref{eq:puncture_initial_time}, using $M_0=1$ and a fitted time translation $C\simeq-9.62$.
    }
    \label{fig:1D_to_ef_coord_vary}
\end{figure*}

We first examine the learned coordinates, shown
in Fig.~\ref{fig:1D_to_ef_coord_vary} at fixed input angles $(\theta,\phi)=(\pi/2,\pi)$. Along the curves shown, the output angles remain close to the input angles, with maximum absolute deviations of $3.1\times10^{-4}$ in $\theta'$
and $6.0\times10^{-4}$ in $\phi'$. This is consistent with the rotational symmetry of the target metric. We therefore focus on the radial and time maps. The radial map has a direct geometric interpretation.
For an angular-coordinate-preserving map, spherical symmetry should give
\begin{equation}
    r'(t,r)
    =\sqrt{g^{\rm NR}_{\theta\theta}(t,r)}
    =r\sqrt{\gamma_T(t,r)}.
    \label{eq:puncture_areal_radius}
\end{equation}
The learned radius in panel~(d) agrees with this
relation to a maximum absolute difference of
$5.9\times10^{-3}$ on the initial slice and
at most $1.5\times10^{-4}$ on the five later time slices shown. For the initial isotropic Schwarzschild data,
Eq.~\eqref{eq:puncture_areal_radius} reduces to
\begin{equation}
    r'(0,r)=
    r\left(1+\frac{M_0}{2r}\right)^2,
    \label{eq:puncture_initial_radius}
\end{equation}
where $M_0=1$ is the mass parameter of the initial
data, distinguished here from the mass inferred from the apparent horizon finder.
This analytical relation is shown by the dashed
black curve in panel~(d), which closely follows
the learned initial radial map.

\begin{figure}
    \centering
    \includegraphics[width=0.5\textwidth]
        {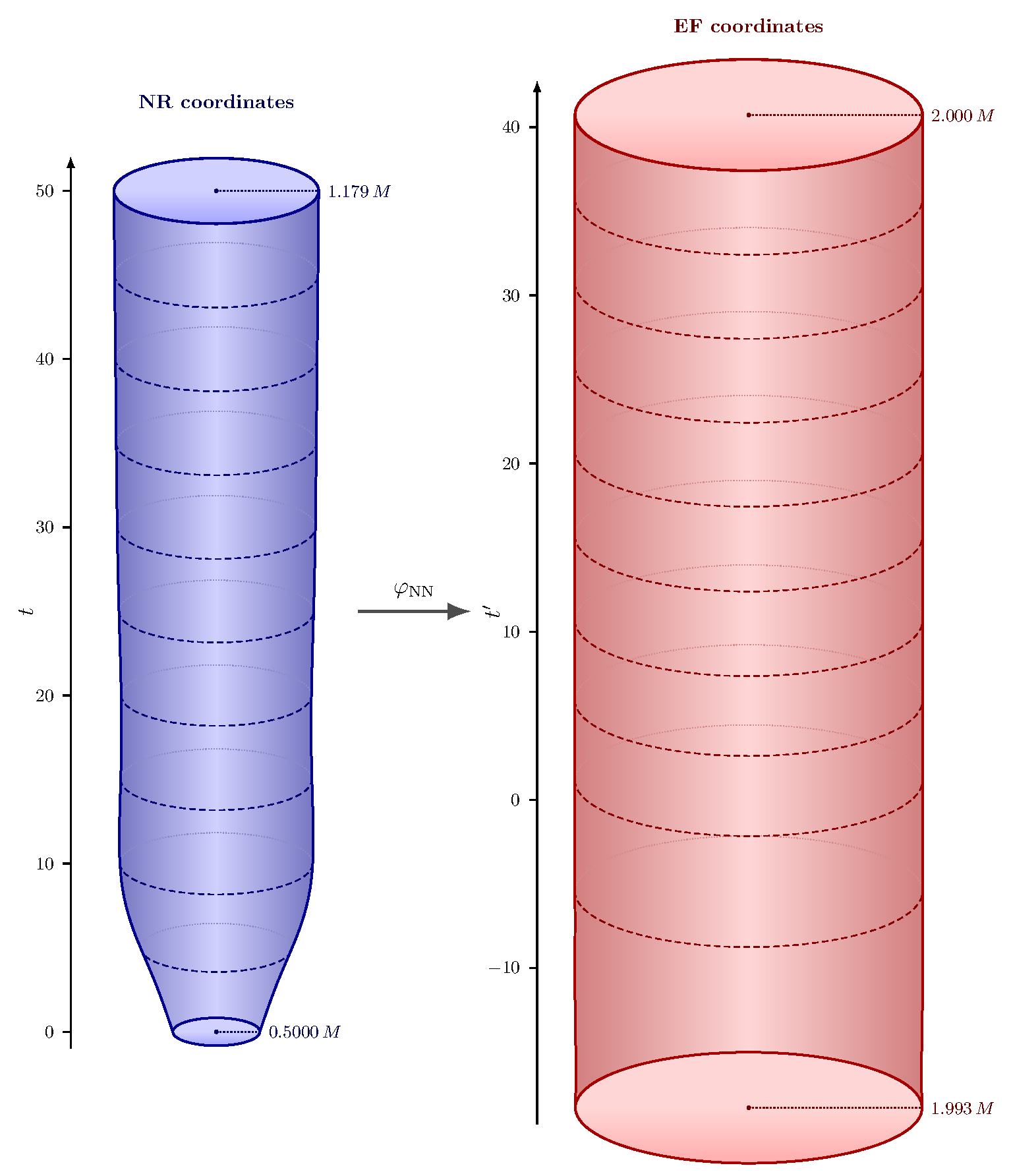}
    \caption{
    Learned mapping of the apparent horizon.
    The blue tube represents equatorial horizon
    rings in numerical coordinates, whose radius
    grows from approximately $0.50M$ to $1.18M$
    over $0\leq t\leq50M$.
    The learned map carries these rings to the
    red tube in \ac{ef} coordinates, where their
    areal radius remains approximately $r'=2M$.
    }
    \label{fig:ahf}
\end{figure}

Panel~(c) shows that, along the fixed input radii
plotted, the learned areal radius generally
decreases during the evolution.
Meanwhile, the apparent-horizon radius in the numerical grid grows from approximately $0.50M$
to $1.18M$, while its learned areal radius remains close to $r'=2M$.
These observations describe the changing relationship between numerical radius and mapped areal radius. The slower late-time variation is consistent with the numerical gauge approaching a
stationary configuration.
Figure~\ref{fig:ahf} illustrates the apparent horizon mapping directly.

Panel~(a) shows the learned time coordinate $t'$ as a function of numerical time at several fixed
numerical radii. At each radius, $t'$ increases approximately
linearly with $t$. The separation between the curves shows that
points at the same numerical time but different radii correspond to different \ac{ef} times. At late times, the curves are nearly parallel,
so these differences remain approximately constant. The learned radial coordinate $r'(t,r)$ also changes little with time at fixed numerical radius during this stage. Together, these observations suggest that later numerical slices approach the same shape in \ac{ef} coordinates and differ mainly by a translation in target time.

The initial slice requires separate consideration. This slice has a constant Schwarzschild time
$T$, and the advanced \ac{ef} time satisfies
\begin{equation}
    t'=T+r_*,
    \qquad
    r_*=r'+2M_0
    \ln\left|\frac{r'}{2M_0}-1\right|.
\end{equation}
Because the tortoise coordinate $r_*$ depends
on radius, $t'$ varies across this slice.
Outside the horizon, where $r'>2M_0$, the
initial time map is therefore
\begin{equation}
    t'(0,r)=
    r'(0,r)+2M_0
    \ln\left[\frac{r'(0,r)}{2M_0}-1\right]+C,
    \label{eq:puncture_initial_time}
\end{equation}
with $r'(0,r)$ given by
Eq.~\eqref{eq:puncture_initial_radius}. The constant $C$ accounts for the arbitrary choice of time origin. The dashed black curve in panel~(b) shows this analytical relation, with $C\simeq-9.62$ fitted over $1\leq r\leq10$. It agrees closely with the learned initial time map away from the horizon.

As the initial slice approaches the horizon, $r_*\to-\infty$ and hence $t'\to-\infty$; the network does not fully resolve this steep initial time dependence near the horizon. At the innermost sampled point discussed here, $r=0.505$, it predicts $r'=1.9965$ which is slightly inside the target horizon. These discrepancies accompany the larger initial near-horizon metric residuals described later in this section. Later numerical slices cross the future horizon at finite advanced \ac{ef} time.
Their learned $t'(r)$ curves consequently remain finite near the apparent horizon. The contrast between the initial and later
curves therefore reflects the evolution of the numerical slicing.

We next assess how accurately the learned map transforms the metric.
Following the notation of
Eq.~\eqref{eq:jacobian}, let
$A^\alpha{}_\mu=\partial x'^\alpha/\partial x^\mu$
be the forward Jacobian and $J=A^{-1}$ the inverse
Jacobian.
The transformed numerical metric is
\begin{equation}
    g'_{\alpha\beta}(x')
    =
    J^\mu{}_\alpha
    J^\nu{}_\beta
    g^{\rm NR}_{\mu\nu}(x),
    \qquad x'=\varphi_{\rm NN}(x).
\end{equation}
We compare this metric with the target \ac{ef} metric
evaluated at the same learned coordinates $x'$.

\begin{figure*}
    \centering
    \includegraphics[width=\textwidth]
        {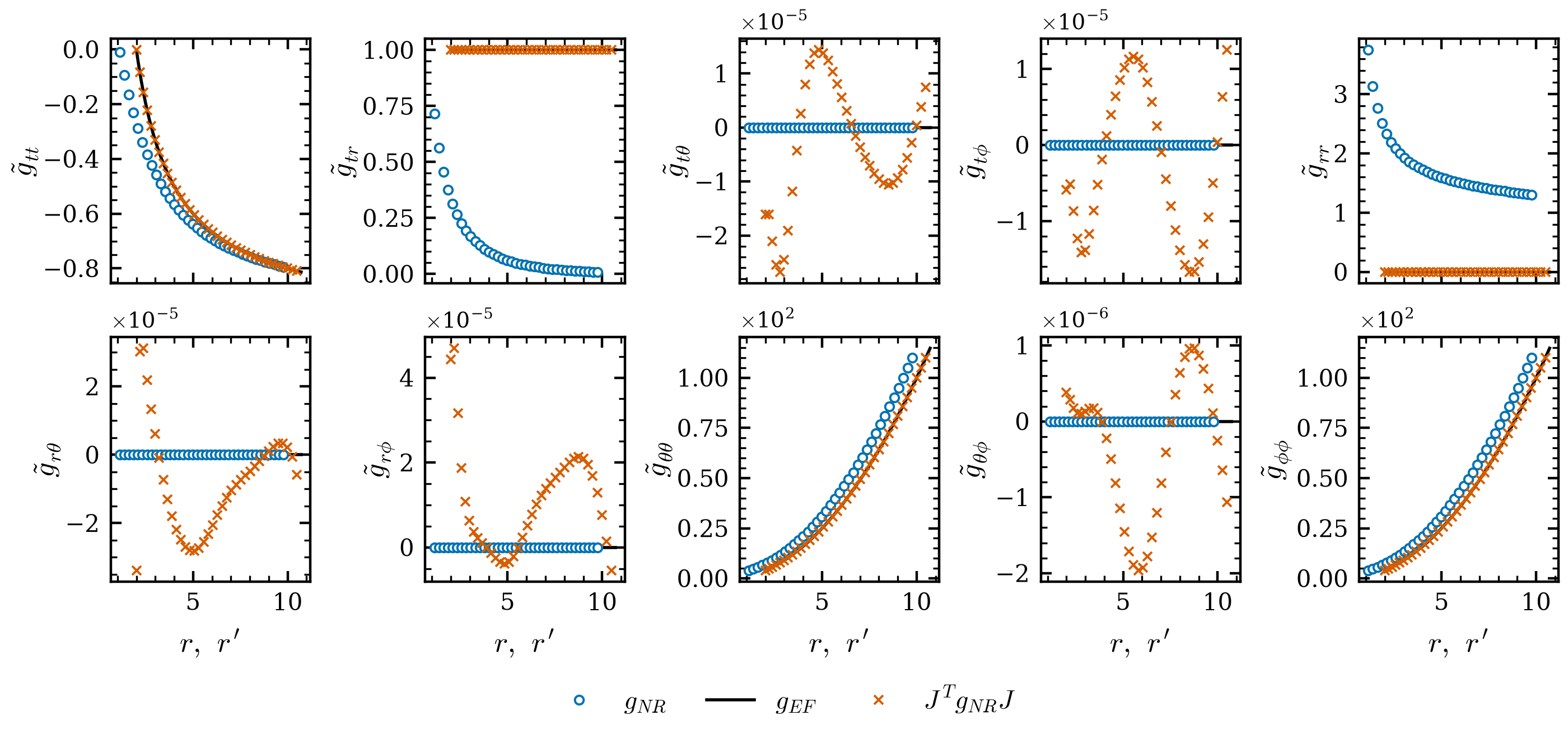}
    \caption{The ten independent metric components for the non-spinning
 puncture at input time $t=25M$. They are evaluated at fixed input angles
 $(\theta,\phi)=(\pi/2,\pi)$ for radii outside the apparent horizon,
 $r_{\rm AH}(t)\leq r\leq10M$. Blue circles show the numerical metric
 $g_{\rm NR}$ against the input radius $r$. The black line shows the
 target ingoing \ac{ef} metric $g_{\rm EF}$, and vermilion crosses show
 the pulled-back metric $J^{T}g_{\rm NR}J$ of
 Eq.~\eqref{eq:metric_transform2}. Both are plotted against the mapped
 radius $r'$. Here $\tilde{g}_{\mu\nu}$ denotes the corresponding component in
 each representation. The five off-diagonal components that vanish in the target metric are shown on magnified scales to display the
small residuals.}
    \label{fig:1D_to_ef_components}
\end{figure*}

Figure~\ref{fig:1D_to_ef_components} shows this
comparison along the equatorial line
$(\theta,\phi)=(\pi/2,\pi)$ of the input slice
$t=25$. Before the transformation, the numerical metric components (shown in circles) differ substantially from the target components (solid line). For example, the numerical $g_{rr}$ decreases from approximately $3.7$ near the horizon to $1.3$ at $r=10$, whereas the target $g^{\rm EF}_{r'r'}$ vanishes. However, after the transformation, the transformed numerical metric components (crosses) closely match the target \ac{ef} components. On this time slice, the maximum absolute residuals are $7.6\times10^{-4}$ in the $r'r'$ component, $4.7\times10^{-4}$ in the $t'r'$ component, and $6.8\times10^{-5}$ in the $t't'$ component. The maximum residual in each angular diagonal component is approximately $3.7\times10^{-4}$.
The other five independent off-diagonal components vanish in the target metric; their transformed residuals are $\mathcal{O}(10^{-5})$. 

Even though the training loss reaches approximately $\mathcal{O}(10^{-8})$, the metric agreement varies across the domain. The larger residuals near the horizon on the initial slice are consistent with the difficulty in learning the initial time map discussed above. Along the initial equatorial line, we found the absolute $r'r'$ residual reaches $1.5\times10^{-2}$ at $r\simeq0.53$, while the angular diagonal residuals reach $2.4\times10^{-2}$ at $r\simeq0.63$. These local discrepancies are not apparent from the training loss alone, which averages the squared residuals over metric components
and sampled points. Appendix~\ref{app:1Dpunc} shows how the metric-matching loss varies with radius and numerical time.

\subsubsection{Single Spinning Puncture}
\label{sec:results.num.spin}
\begin{figure}
 \centering
\includegraphics[width=\columnwidth]{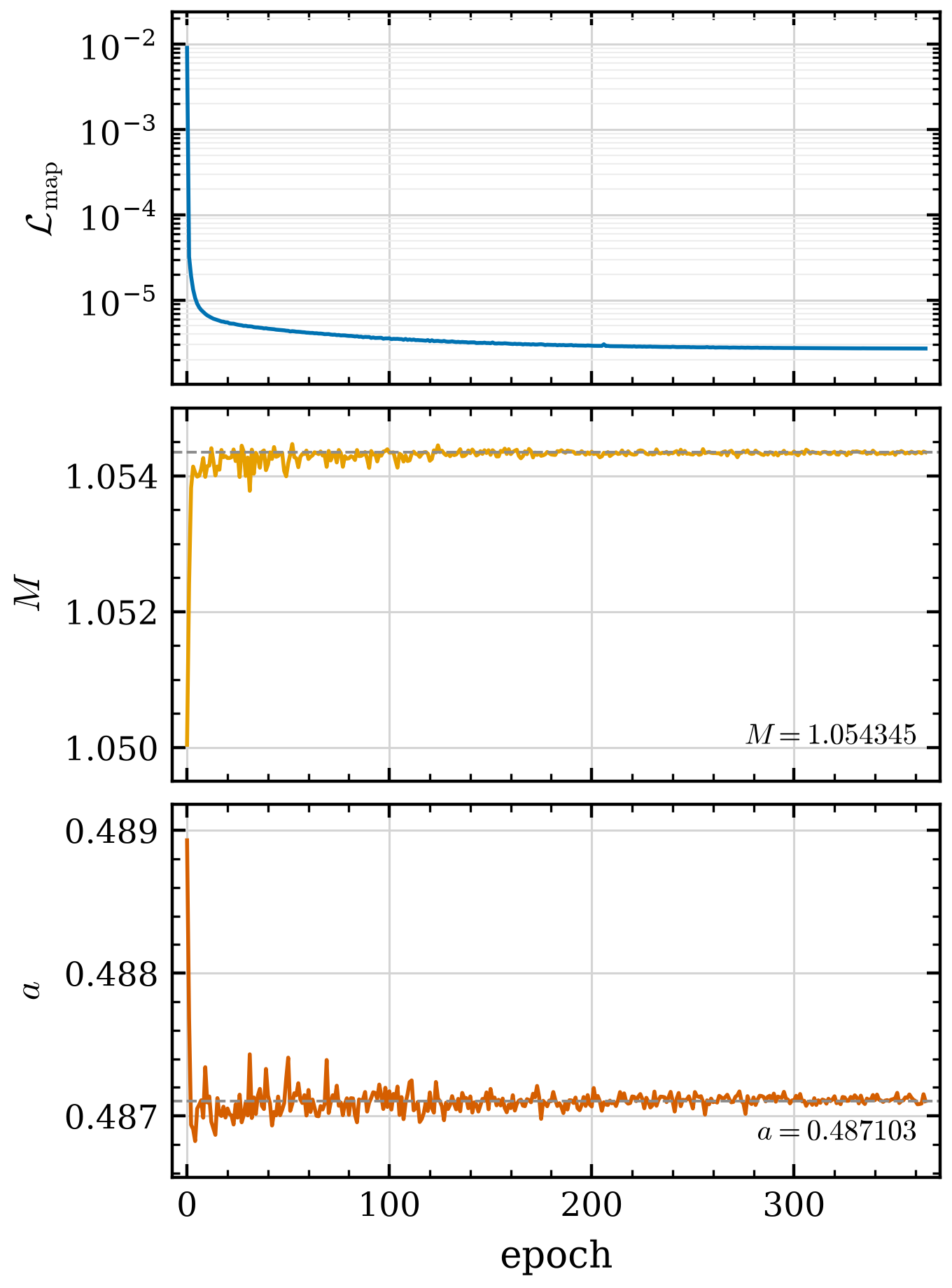}
 \caption{Training loss $\mathcal{L}_{\mathrm{map}}$ for the map from the perturbed spinning puncture to Cartesian \ac{ks} coordinates (top), together with the fitted mass $M$ (middle) and spin parameter $a$ (bottom), shown as functions of epoch. The loss, plotted on a logarithmic scale, decreases from $\mathcal{O}(10^{-2})$ at initialization to $\mathcal{O}(10^{-6})$ by approximately 350 epochs. The dashed horizontal lines mark the final fitted values, $M=1.054345$ and $a=0.487103$.}
 \label{fig:spin_loss}
 \end{figure}

As a final experiment, we apply this technique to map the perturbed single spinning \ac{bh} described in Sec.~\ref{sec:setup.kerr} to Cartesian \ac{ks} coordinates (Appendix~\ref{app:metrics}). We treat the mass $M$ and spin parameter $a$ of the target metric as learnable parameters, initialized at $M=1$ and $a=0.5$. We set these initial guesses close to the simulation parameters of the initial data. As shown in Fig.~\ref{fig:spin_loss}, the mapping loss decreases rapidly during the initial epochs and then more gradually, approaching a plateau of $\mathcal{O}(10^{-6})$ by approximately 350 epochs. The fitted parameters settle near $M=1.0543$ and $a=0.4871$, within $0.21\%$
and $0.31\%$, respectively, of the settled apparent-horizon values
$M_{\rm H}=1.0566$ and $a=J_{\rm H}/M_{\rm H}=0.4856$. As discussed earlier, the simulated spacetime is perturbed and cannot be mapped exactly to stationary Kerr by a coordinate transformation. By jointly optimizing the coordinate map and the background parameters using the loss in Eq.~\ref{eq:loss}, the network selects an approximate identification between the numerical spacetime and a Kerr background~\cite{Stewart:1974uz}. The residual mismatch contains physical deviations from Kerr together with numerical and fitting errors. In Appendix~\ref{app:massspin}, we further investigate how the choice of training domain affects the fitted parameters. 

To further assess how well the learned map preserves the radiative content of the perturbation, we compute $\psi_m$ (Eq.~\ref{eq:m-projection}) and its background-coordinate time derivative $\partial_{t'}\psi_m$ on slices of constant \ac{ks} time, using the \ac{nn} map and a null tetrad associated with the Kerr background (see Sec.~\ref{sec:setup.teuk} for details). These quantities provide initial data for the Teukolsky equation \eqref{eq:teuk_ks}. The aim is to compare Teukolsky evolutions initialized at different \ac{ks} times with the \ac{nr} waveform to investigate when a linear perturbation theory becomes an adequate description of the spacetime.

\begin{figure}
 \centering
\includegraphics[width=\columnwidth]{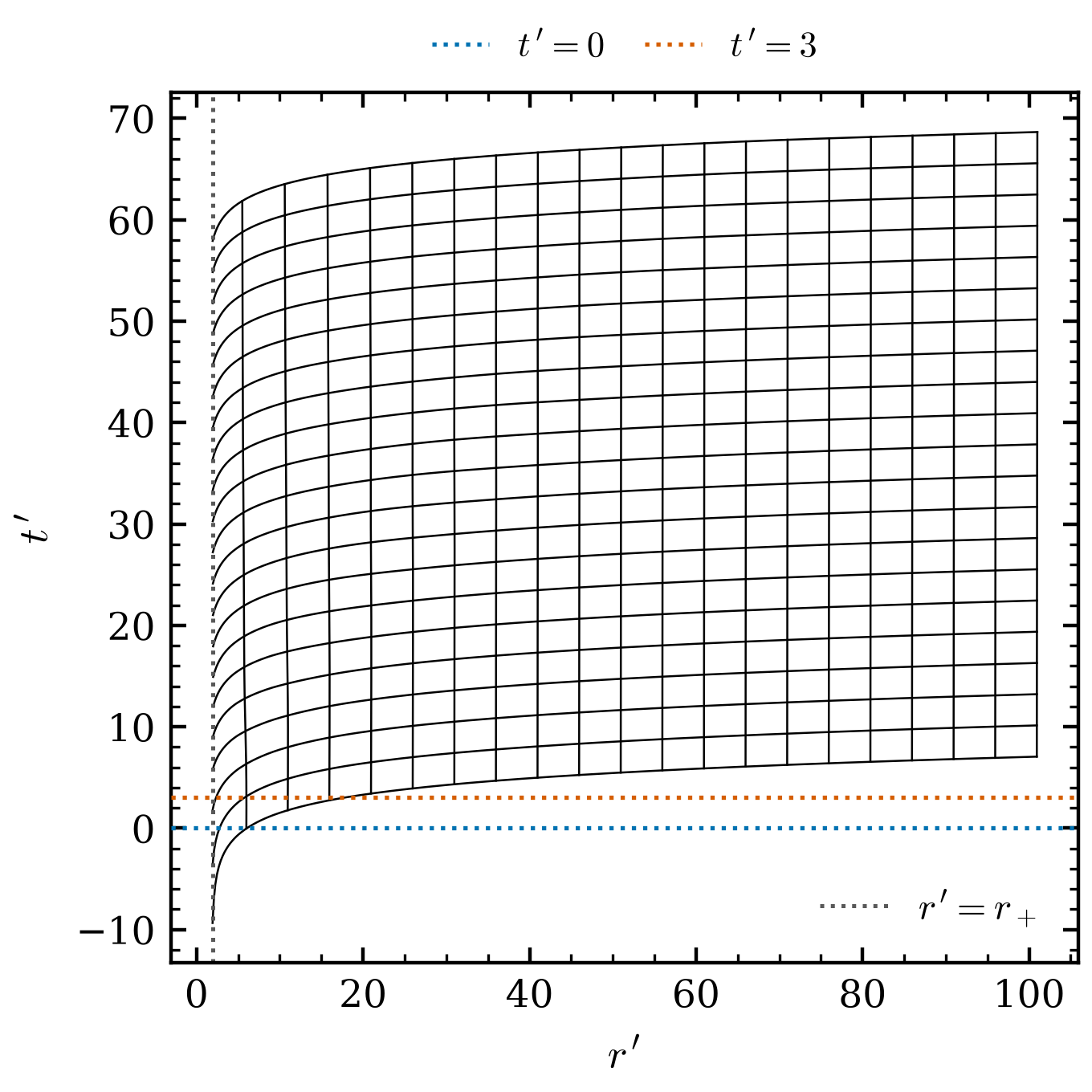} 
 \caption{Numerical coordinate grid mapped to Kerr--Schild
coordinates $(t',r')$ at fixed numerical angles
$\theta=\pi/2$ and $\varphi=\pi$. The roughly horizontal
black curves represent constant numerical time, while
the roughly vertical black curves represent constant
numerical radius. Colored horizontal dotted lines
indicate the extraction slices $t'=0,3$ for the Teukolsky evolution, while the vertical gray dotted line marks the background Kerr
horizon, $r'=r_+$.}
  \label{fig:spin_to_ks_coord}
 \end{figure}
Fig.~\ref{fig:spin_to_ks_coord} shows the numerical coordinate
grid mapped to \ac{ks} coordinates by the \ac{nn}. The
horizontal solid black curves correspond to constant
$t_{\rm NR}$, while the vertical solid black
curves correspond to constant $r_{\rm NR}$. The colored
horizontal dotted lines mark the constant \ac{ks}
times at which we choose to construct the Teukolsky
initial data. At larger radii, each constant-$t'$ surface samples earlier numerical times, so the finite training interval limits its radial coverage. At $t'=0$ the initial data for the Teukolsky evolution extend only to
$r' \approx 6.0$, and at $t'=3$ to $r' \approx 17.5$. Beyond these radii, NR
data are missing either at some sampled angle or at some time in the stencil
used to compute $\partial_{t'}\psi_m$. Therefore, to initialize the Teukolsky evolution on the full solver grid, we set both $\psi_m$ and $\partial_{t'}\psi_m$ to zero
at grid points with $r'\leq r_+$ and on radial shells
without complete extraction data up to the outer radial boundary of the solver at $r_{\rm max} = 400$ M. The horizon-interior
values do not enter the exterior evolution because the
solver uses one-sided radial stencils near the horizon,
consistent with excision. At larger radii, zero filling
supplies initial data without extrapolating the \ac{nn} map beyond its training domain. However, this truncates the initial perturbation. Despite that we do not identify a numerical instability associated with the cutoff in the extracted waveforms. It may indicate that the cutoff-induced error is small compared with the physical signal, although we have not quantified its contribution. 

\begin{figure*}
 \centering
\includegraphics[width=\textwidth]{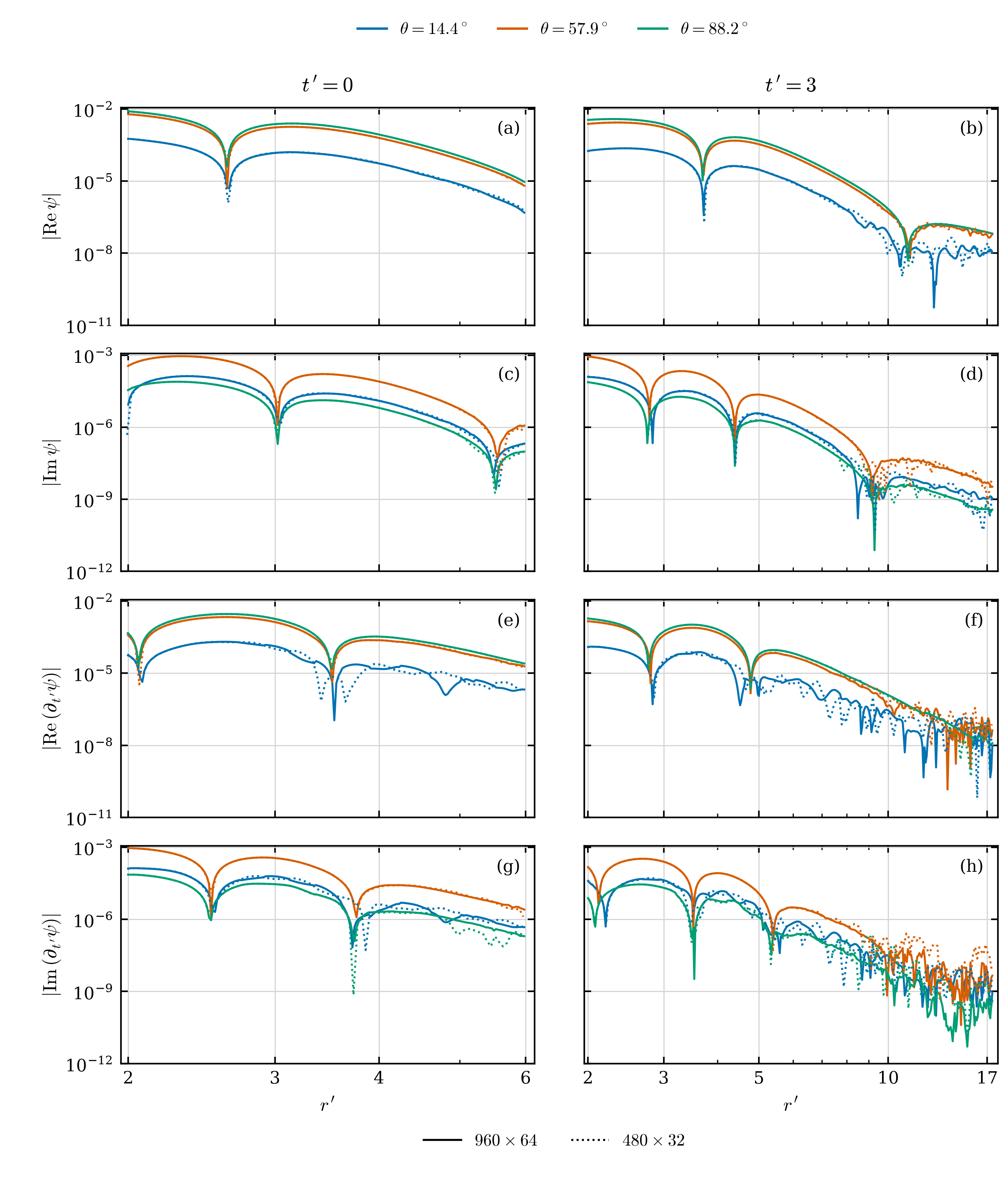}
\caption{Absolute values of the real and imaginary
components of the $m=0$ Teukolsky field $\psi$ and
its time derivative $\partial_{t'}\psi$, shown as
functions of Kerr--Schild radius $r'$ on logarithmic
axes. Columns correspond to $t'=0$ (left) and $t'=3$
(right). From top to bottom, rows show
$|\mathrm{Re}\,\psi|$, $|\mathrm{Im}\,\psi|$,
$|\mathrm{Re}\,\partial_{t'}\psi|$, and
$|\mathrm{Im}\,\partial_{t'}\psi|$.
Colors distinguish polar angles, while solid and
dotted curves correspond to resolutions
$(N_r,N_\mu)=(960,64)$ and $(480,32)$, respectively,
where $\mu=\cos\theta$. Only radial points with valid
initial data are shown; the remaining radial points
up to the solver's outer boundary $r_{\max}$ are
initialized to zero.}
\label{fig:init_pert}
 \end{figure*}

Fig.~\ref{fig:init_pert} shows the initial $m=0$ perturbation and its time derivative at $t'=0$ and $t'=3$ for several polar angles. The profiles vary with both radius and angle, with the largest amplitudes concentrated near the strong-field region. The later-time perturbation extends to larger radius because the corresponding constant $t'$ surface remains within the numerical training interval over a larger radial domain. We notice that the two resolutions agree closely where the perturbation is larger in magnitude and smooth. At larger radii, where the perturbation is much weaker, the profiles become irregular and the
agreement deteriorates, particularly for the time derivative. This behavior suggests that numerical noise limits the accuracy of the reconstructed
initial data in these regions. The decay of the signal toward the outer edge
of the available data motivates setting $\psi_m$ and $\partial_{t'}\psi_m$ to zero beyond that edge, where reconstruction is unavailable.
This assumes that the truncated outer perturbation has a small effect on the evolved waveform. We evolve both these initial data with \texttt{PyTeukolsky}\cite{pyteukolsky} at two numerical resolutions for a duration of $200$ M,
using the fitted mass and spin to specify the Kerr background.

\begin{figure*}
 \centering
\includegraphics[width=\textwidth]{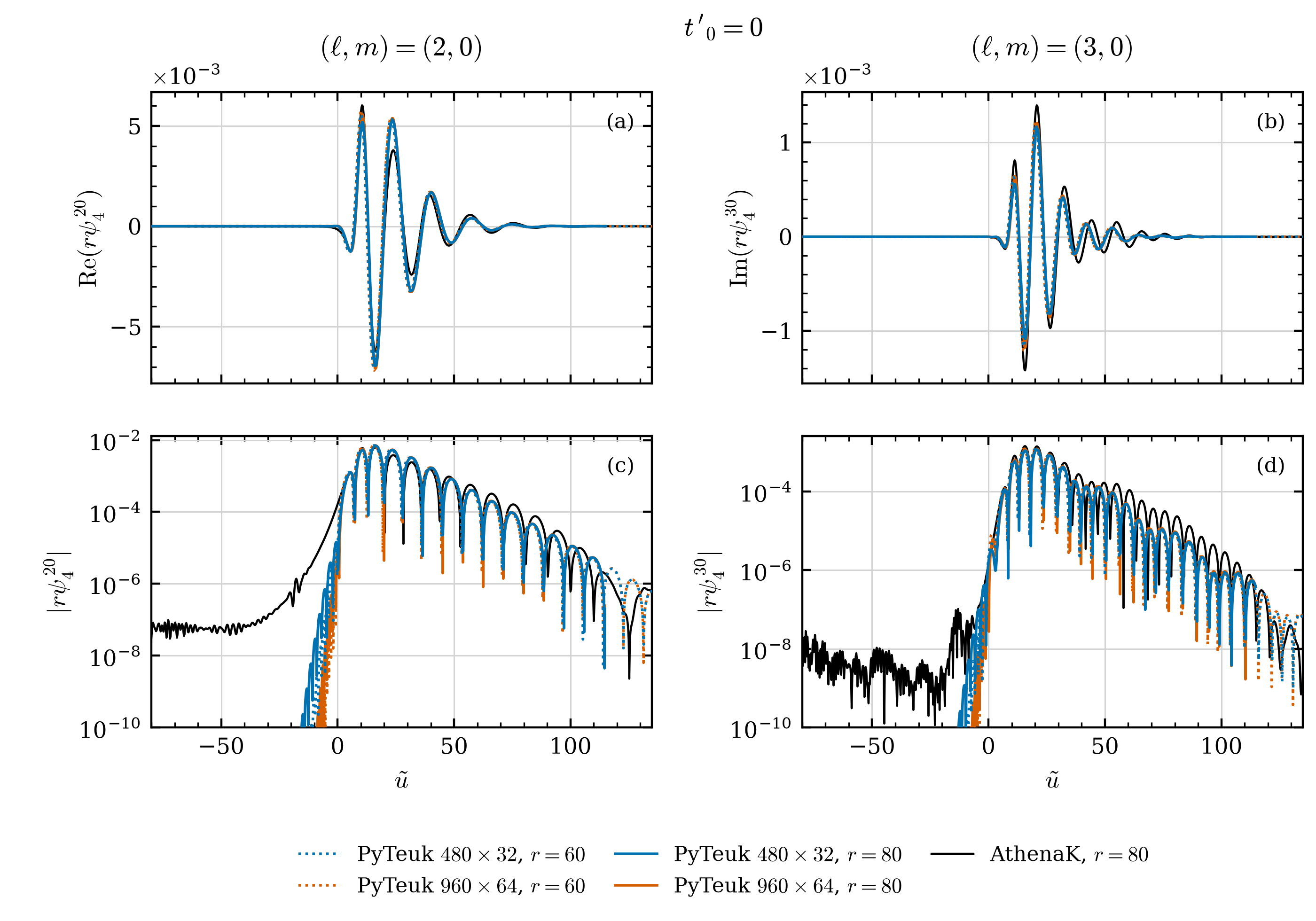}
\caption{Teukolsky waveforms from initial data extracted at $t'_0=0$, compared with \texttt{AthenaK}. The left and right columns show the
$(\ell,m)=(2,0)$ and $(3,0)$ modes, respectively. The top row
shows $\mathrm{Re}(r\psi_4^{20})$ (a) and
$\mathrm{Im}(r\psi_4^{30})$ (b); the bottom row shows the
complex magnitude $|r\psi_4^{\ell 0}|$ on a logarithmic
vertical axis (c, d). Blue and orange curves are
\texttt{PyTeuk} evolutions at resolutions
$(N_r,N_\mu)=(480,32)$ and $(960,64)$, respectively. Dotted
and solid colored curves denote extraction radii $r=60$ and
$80$. The solid black curve shows \texttt{AthenaK} at $r=80$.
The waveforms are compared after tetrad conversion and a
constant time alignment, with aligned time denoted by
$\tilde{u}$.}
\label{fig:spin_wave_t0}
 \end{figure*}

\begin{figure*}
 \centering
\includegraphics[width=\textwidth]{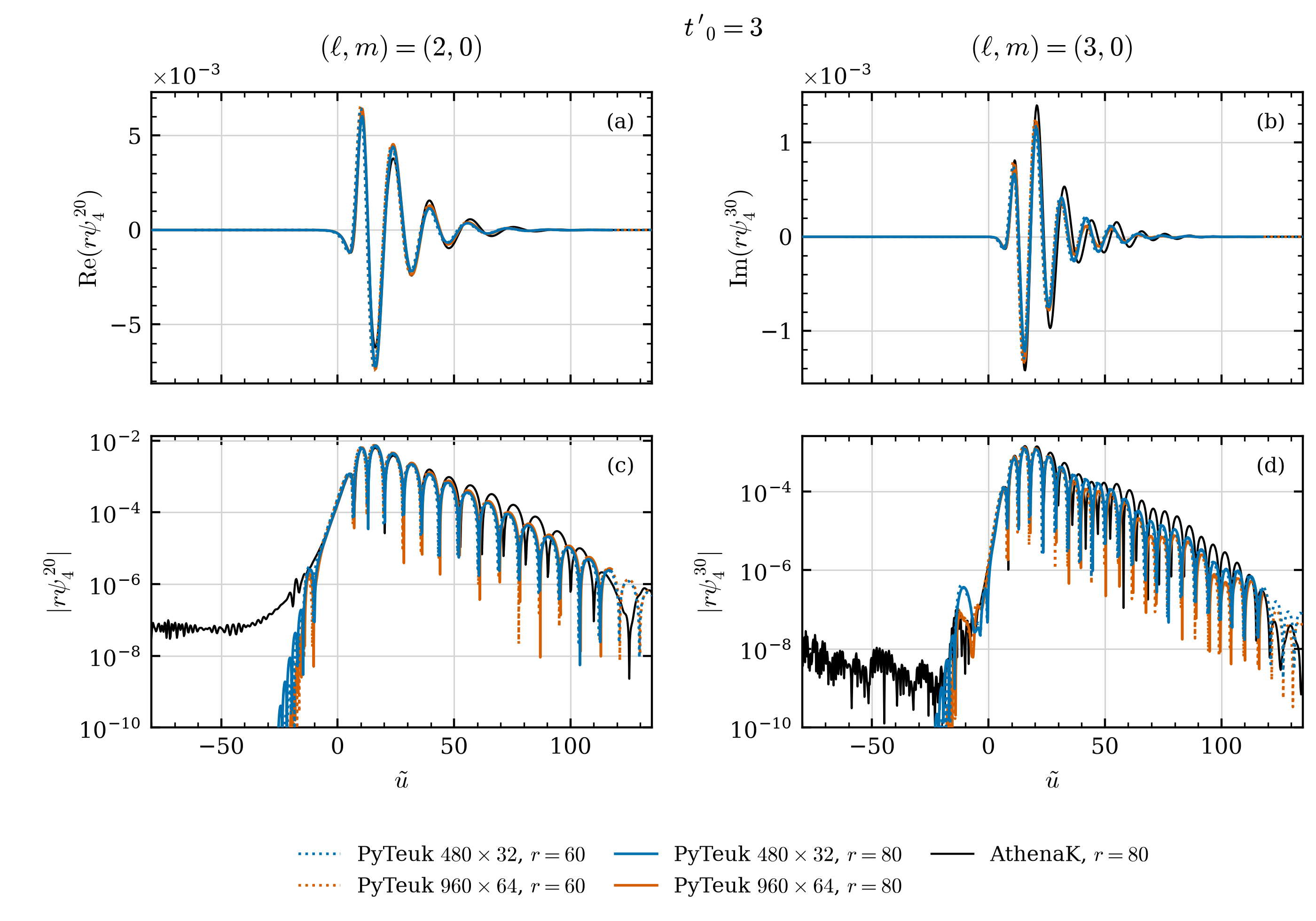}
\caption{Teukolsky waveforms from initial data extracted at $t'_0=3$, compared with
\texttt{AthenaK}. The left and right columns show the
$(\ell,m)=(2,0)$ and $(3,0)$ modes, respectively. The top row
shows $\mathrm{Re}(r\psi_4^{20})$ (a) and
$\mathrm{Im}(r\psi_4^{30})$ (b); the bottom row shows the
complex magnitude $|r\psi_4^{\ell 0}|$ on a logarithmic
vertical axis (c, d). Blue and orange curves are
\texttt{PyTeuk} evolutions at resolutions
$(N_r,N_\mu)=(480,32)$ and $(960,64)$, respectively. Dotted
and solid colored curves denote extraction radii $r=60$ and
$80$. The solid black curve shows \texttt{AthenaK} at $r=80$.
The waveforms are compared after tetrad conversion and a
constant time alignment, with aligned time denoted by
$\tilde{u}$.
}
\label{fig:spin_wave_t3}
 \end{figure*}

Fig.~\ref{fig:spin_wave_t0} compares the Teukolsky
waveforms extracted at $r=60$ and $r=80$, initialized
from the constant-$t'=0$ Cauchy surface, with the
\texttt{AthenaK} waveforms. The perturbative evolution
captures the overall waveform structure in the
$(2,0)$ and $(3,0)$ modes, with good
qualitative agreement in the dominant real $(2,0)$
and imaginary $(3,0)$ components. After time alignment,
the main peaks and troughs closely track the numerical
signal, followed by a similar sequence of damped
oscillations. The results at both extraction radii
and numerical resolutions closely overlap through the main signal. This agreement suggests that the learned map and the chosen background tetrad provide a suitable prescription for constructing Teukolsky initial
data that reproduce the main features of the
numerical waveform. 
For initial data constructed at the later time
$t'=3$ (Fig.~\ref{fig:spin_wave_t3}), the dominant
real $(2,0)$ and imaginary $(3,0)$ components show
slightly closer agreement with \texttt{AthenaK}
through the main oscillations than in the $t'=0$
case. Such an improvement is expected as the
perturbed black hole relaxes toward a stationary
Kerr solution and linear perturbation theory becomes
a more accurate description. The agreement in these
dominant components is maintained at both extraction radii. 

Lastly, we observed that evolutions initialized at even later extraction times developed a late-time numerical instability after the initial physical burst. This instability may be caused in part by large constraint violations in the initial data, arising from the under-resolved smaller puncture, and it limits the extraction times considered here.

\section{Summary}
\label{sec:summary}
In this paper, we present a \ac{nn} method for learning coordinate maps between spacetime metrics. The network takes the source coordinates as input and returns coordinates in a prescribed target gauge. We specify this gauge by fixing the functional form of the target spacetime metric. Automatic differentiation provides the Jacobian needed to transform the source metric, and the training minimizes its mismatch with the target. Each network is fitted to a particular mapping over a specified spacetime domain.

We first tested the method on analytical Schwarzschild and Kerr geometries. In both cases, the network recovered the known coordinate transformations up to time translations and rotations allowed by the symmetries of the target metric. The relative Jacobian errors had root-mean-square
values of $4\times10^{-5}$ and $5\times10^{-5}$
for Schwarzschild and Kerr, respectively.
These tests demonstrate that metric matching
can determine accurate coordinate maps without
assuming their functional form.
They also illustrate that the target metric
fixes the coordinates only up to its isometries.

For the non-spinning numerical puncture, the
learned map relates the evolving moving-puncture
coordinates to a stationary Schwarzschild metric
in ingoing \ac{ef} coordinates.
The radial map recovers the areal radius and
maps the evolving numerical apparent horizon
to an approximately constant target radius.
The time map captures the changing numerical
slicing, including the distinction between the
initial constant-Schwarzschild-time slice and
the later slices that cross the future horizon.
The metric-matching loss reaches approximately
$4.8\times10^{-8}$, with larger local residuals
near the horizon on the initial slice, where
the exact \ac{ef} time varies steeply.
This test shows that the method can account
for substantial gauge dynamics even when the
underlying spacetime is static.

For the spinning puncture, the numerical
spacetime contains physical departures from
Kerr background. Here, metric matching provides an approximate
identification with a Kerr background, while
jointly fitting the background mass and spin to within $0.21\%$ and $0.31\%$ of the late-time apparent-horizon values respectively. Using the learned identification, we constructed curvature perturbations on constant Kerr--Schild time slices and evolved them with the Teukolsky equation. The resulting waveforms reproduce the main features of the full nonlinear waveforms, with close agreement in the dominant real $(2,0)$ and imaginary $(3,0)$ components. Initialization at later time slices gives slightly better agreement through the main oscillations, consistent with the spacetime approaching Kerr. These results show that the method is capable of identifying a Kerr background in a perturbed numerical spacetime and provide initial data for a perturbative evolution that reproduces the dominant features of the full nonlinear waveform.

We note that using this methodology provides a systematic alternative to the coordinate identification used in the original Lazarus approach \cite{Baker:2001sf}. That prescription relied on identifying the numerical time and polar angle with their \ac{bl} counterparts, supplemented by a curvature-based estimate of the radial coordinate and a correction to the azimuthal angle \cite{Campanelli:2005ia}. Here, the identification is instead determined by optimizing an explicit measure of agreement between the metrics over the training domain, while simultaneously fitting the background mass and spin. Additionally, the residual loss will depend on the sampling data and training accuracy. It provides a measure of the remaining metric mismatch after gauge alignment from the \ac{nn} training and help quantify deviations from the Kerr background.

There are several aspects of this technique that require further development. First, the present network architecture does not enforce invertibility explicitly. Building that explicitly in the network's layers would provide a more robust control over the coordinate map. This could be done using a similar approach used in \cite{Celledoni_2023}. 

Second, the loss function could be better adapted to the different scales of the metric components. The current loss averages their squared absolute differences, so components with large coordinate-dependent magnitudes can dominate the optimization. This is especially relevant for angular components at large radius in spherical coordinates. A loss with suitable component scaling and spatial weighting could give more balanced accuracy across the domain.

Third, the curvature extraction uses a tetrad
associated with the fitted Kerr background.
A quasi-Kinnersley construction based on the
numerical curvature could reduce errors due
to tetrad misalignment and improve the
identification of the radiative perturbation,
particularly in the weaker waveform components
\cite{Campanelli:2005ia}. The perturbative application has additional sources of uncertainty.
Reconstructing curvature from interpolation and taking its time derivative introduces numerical noise, which becomes more significant as the signal weakens. The finite training domain also limits the radial extent of the reconstructed initial data, requiring an outer cutoff whose effect on the waveform remains to be quantified.

Nonetheless, this work introduces a novel approach to learning coordinate maps directly from spacetime metric data, with potential applications to comparing numerical simulations in different gauges and identifying perturbations of analytical
backgrounds. A natural extension is the comparison of simulations produced by different numerical
relativity codes. Learning a map between their coordinate
descriptions could make comparisons of local spacetime quantities more direct. For simulations of the same physical system, this would help distinguish differences due
to gauge choice from differences due to numerical error.

Extension of this method could also be applied to binary black-hole simulations. Maps between numerical descriptions of the same binary could support comparisons post merger. An identification with a Kerr background could provide initial data for perturbative evolution as in earlier hybrid numerical--perturbative
approaches~\cite{Price:1994pm,Baker:2001sf}, and help assess when linear theory adequately describes the remnant, a question that remains open in ringdown modeling~\cite{PhysRevD.97.104065,Giesler:2019uxc,Baibhav:2023clw,Mitman:2022qdl,Cheung:2022rbm}.

\begin{acknowledgments}
We would like to thank Abhay Ashtekar, Hengrui Zhu and Vaishak Prasad for helpful comments and discussions.
ES acknowledges support from The Charles E. Kaufman Foundation of The Pittsburgh Foundation. SB acknowledges funding from the EU Horizon under ERC Consolidator Grant no. InspiReM-101043372 and from the Deutsche Forschungsgemeinschaft (DFG) project ``GROOVHY'' (BE 6301/5-1 Projektnummer: 523180871). 
This research used resources of the National Energy Research Scientific
Computing Center (NERSC), a Department of Energy User Facility using NERSC award ERCAP0039050. Additional computations were performed on the Roar supercomputer at the
Penn State Institute for Computational and Data Sciences (ICDS). 
Claude AI was used to help with coding and plotting scripts.
\end{acknowledgments}

\appendix
\section{1+1D puncture}
\label{app:1Dpunc}
\begin{figure*}
\includegraphics[width=\columnwidth]
{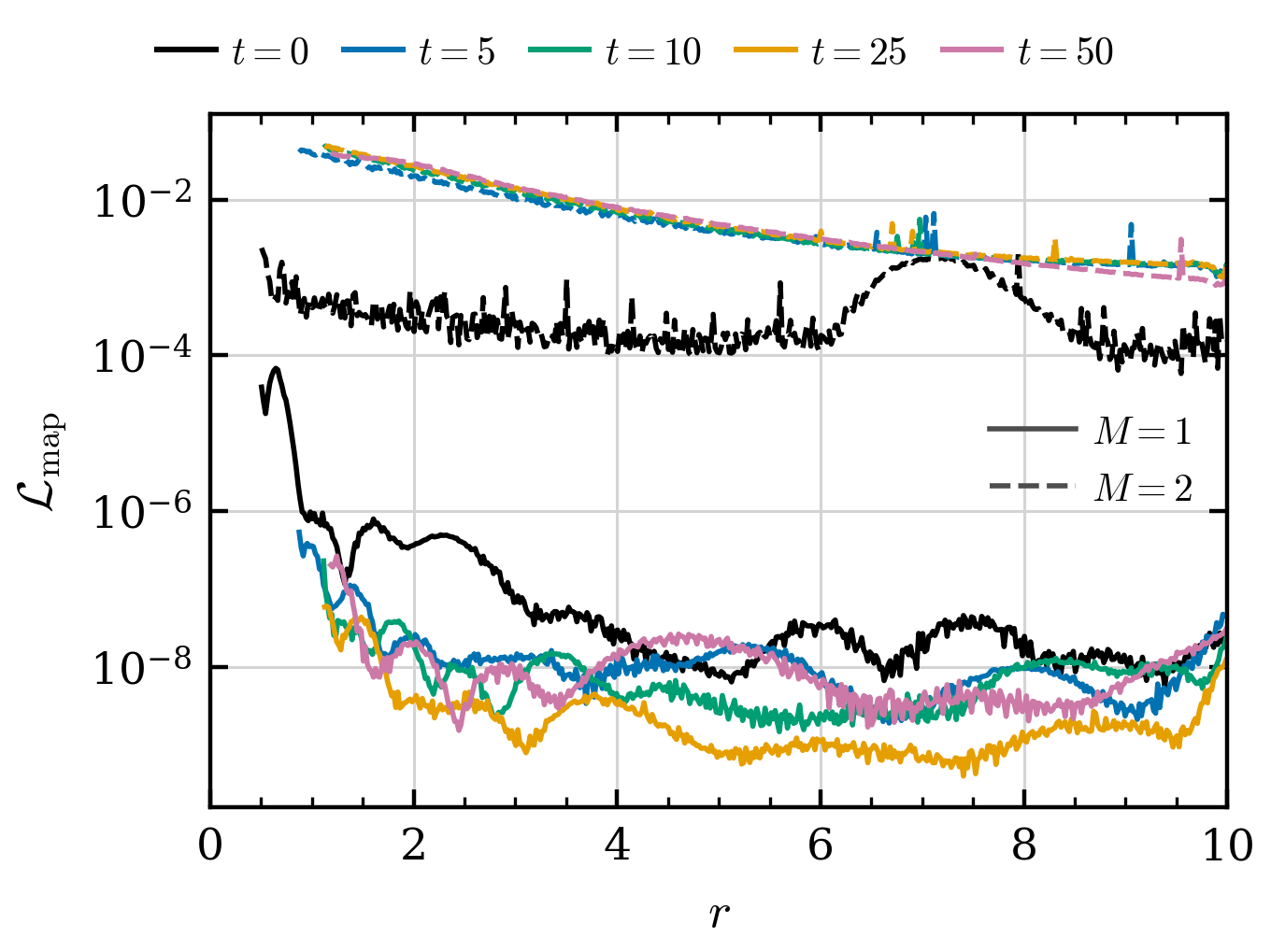} 
 \caption{Metric-matching loss as a function of the input radius
 $r$ at \ac{nr} times $t=0,5,10,25,50$ (colours).
 Solid curves correspond to the correct target mass $M=1$,
 and dashed curves to the incorrect target mass $M=2$.
 The loss is averaged over the sampled angular coordinates
 and all metric components at each radius.
 The incorrect target mass yields substantially larger residuals,
 while the correct-mass fit shows increased errors at small
 radii, most prominently at $t=0$.}
  \label{fig:1D_to_ef_loss_r}
 \end{figure*}

We attempted to map the $1+1$D puncture data to ingoing
\ac{ef} coordinates with an incorrect target mass, $M=2$. Figure~\ref{fig:1D_to_ef_loss_r} shows the radial dependence
of the metric-matching loss at several \ac{nr} time slices,
comparing the correct target mass $M=1$ with $M=2$.
The incorrect-mass model produces substantially larger residuals since a coordinate transformation cannot remove the physical difference
between Schwarzschild geometries with different masses.
For the correct-mass model, the residuals generally increase
toward smaller radii, particularly on the initial $t=0$ slice.
This agrees with the observation in Sec.~\ref{sec:results.num.1d} that fitting the initial slice to \ac{ef} is more difficult near the horizon.

\begin{figure*}
\includegraphics[width=\columnwidth]
{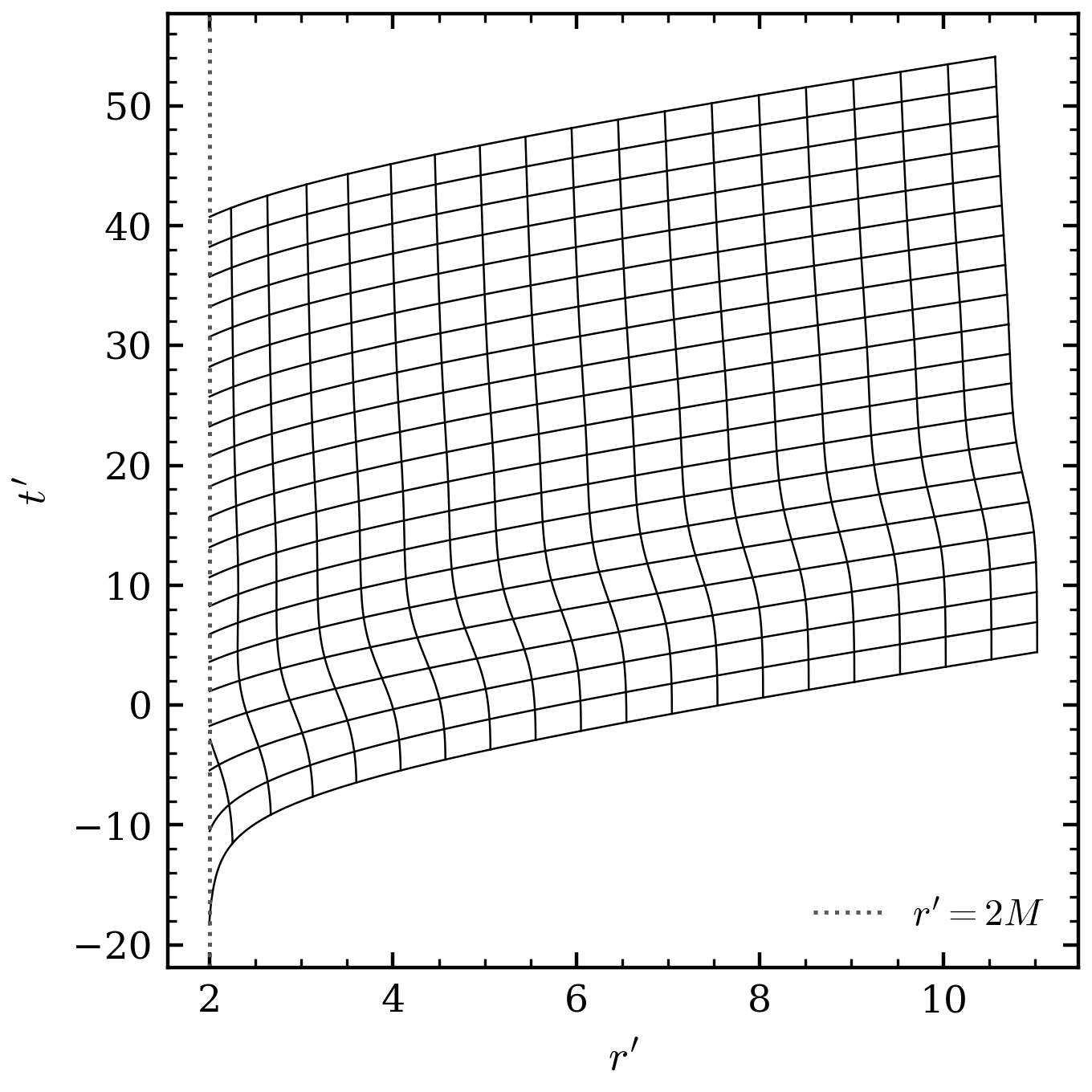} 
 \caption{Numerical puncture coordinate lines mapped to the
ingoing \ac{ef} chart by the learned
transformation with target mass $M=1$.
Black curves represent constant-$t$ slices and constant-$r$
lines, sampled at intervals $\Delta t=2.5$ and $\Delta r=0.5$,
respectively, with $(\theta,\phi)=(\pi/2,\pi)$ held fixed.
The displayed region covers $0\leq t\leq50$ and
$r_{\rm AH}(t)\leq r\leq10$.
The dotted line denotes the target horizon, $r'=2M$.}
  \label{fig:1D_to_ef_coord}
 \end{figure*}

To illustrate the learned coordinate transformation,
Fig.~\ref{fig:1D_to_ef_coord} shows the numerical puncture
coordinate grid mapped into the target
\ac{ef} chart for $M=1$.
Along each constant-$t$ slice, the transformed time $t'$
increases with the transformed radius $r'$. Thus, a single
numerical time slice spans a range of \ac{ef} times, with the strongest curvature near the inner boundary on the earliest slices.

The mapped constant-$r$ lines also bend appreciably at early
times, showing that a fixed numerical radius does not
correspond to a fixed target radius during this stage of
the evolution. At later times, these lines become nearly
vertical. This pattern
suggests that the coordinate transformation approaches
a more nearly time-independent radial map at late times.
The inner edge of the mapped grid remains close to
$r'=2M$, consistent with the numerical apparent horizon
being mapped to the horizon of the target geometry.

\section{Learning mass and spin}
\label{app:massspin}
\begin{figure*}
  \centering
\includegraphics[width=\columnwidth]{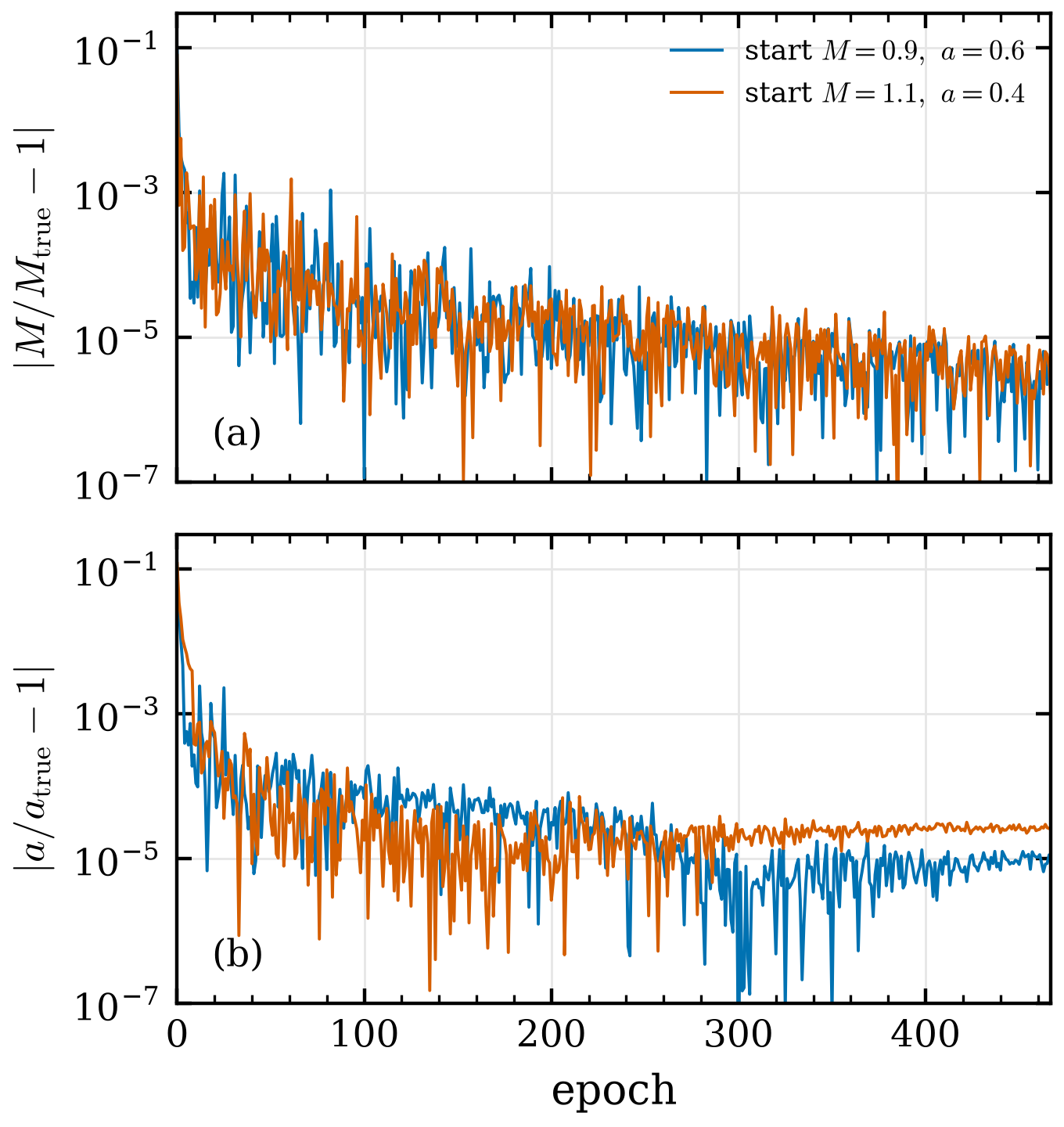}
  \caption{Relative error of the learned (a) mass and (b) spin parameter for
  the analytic Kerr map (true values $M=1$, $a=0.5$), from two initial guesses.}
  \label{fig:mafit_analytic}
\end{figure*}

To test whether this methodology is capable of recovering the correct mass and spin parameters, we repeat the analytical test (see Sec.~\ref{sec:results.analyt.kerr}) of identifying coordinate transformation from \ac{bl} coordinates to ingoing Kerr coordinates with $M$ and $a$ as learnable parameters. The true values are $M=1$ and $a=0.5$. We train two networks with initial guesses as $(M,a)=(0.9,0.6)$ and
$(1.1,0.4)$. Figure~\ref{fig:mafit_analytic} shows that both runs converge to the same, correct values. The final relative errors are $\mathcal{O}(10^{-6})$ in $M$
and $\mathcal{O}(10^{-5})$ in $a$, independent of the starting point. We found that the learned coordinate map agrees with the exact one as closely as a map trained with $M$ and $a$ held fixed at their true values.

\begin{figure*}
  \centering
  \includegraphics[width=\linewidth]{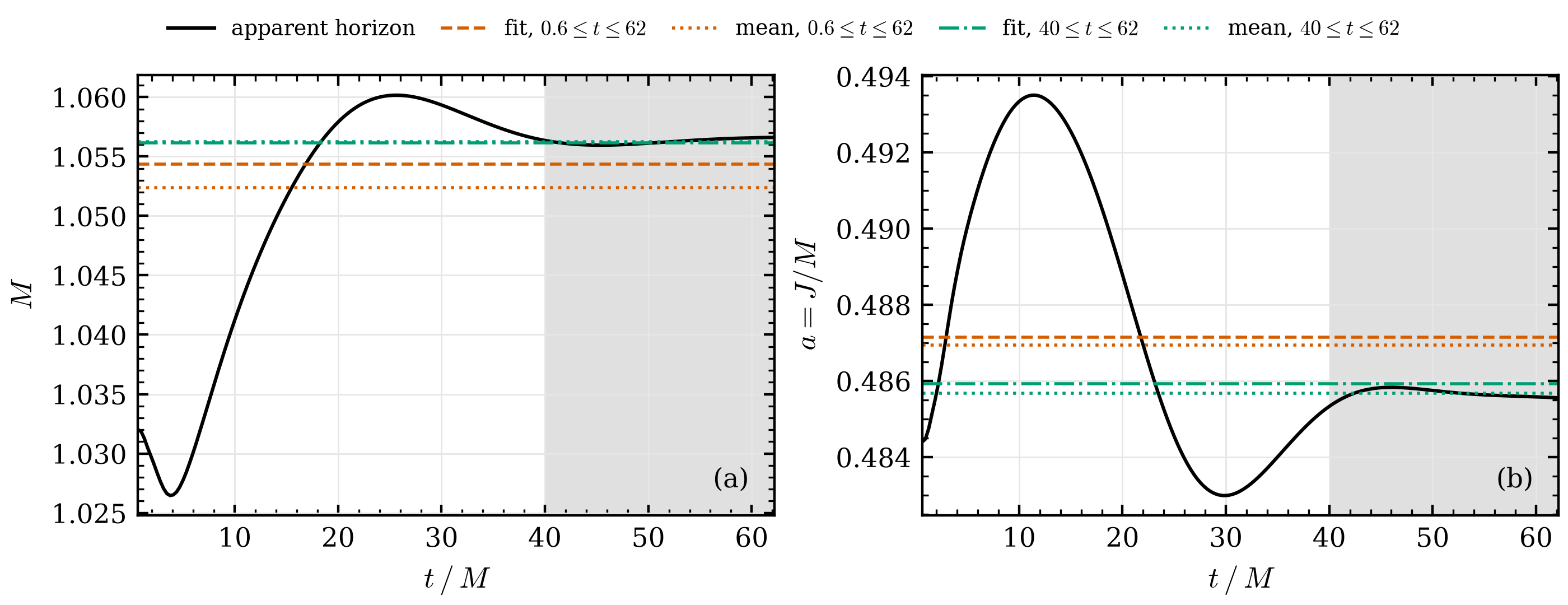}
  \caption{Apparent-horizon (a) mass and (b) spin (solid) over the time range of the training data used in Sec.~\ref{sec:results.num.spin}, compared with the Kerr parameters learned from the full window $0.6\le t\le62$ (dashed) and from $40\le t\le62$ only (dash-dotted, shaded). Dotted lines give the horizon values averaged over each window, in the matching colour.}
  \label{fig:mafit_horizon}
\end{figure*}

In case of the numerical simulation of perturbed Kerr \ac{bh} in Sec.~\ref{sec:results.num.spin}, the fit projects the \ac{nr} solution onto the Kerr family, and tries to find a Kerr manifold that minimizes the deviation from the physical metric the most. Quasi-local horizon quantities and the parameters of a single Kerr background are expected to agree when the spacetime is stationary. Figure~\ref{fig:mafit_horizon} shows that this is not the case for most of the training window. The apparent horizon is still changing from the initial data (solid curves): $M_{\rm H}$ ranges from $1.027$ to $1.060$ [panel (a)] and $a_{\rm H}$ from $0.483$ to $0.494$ [panel (b)] before settling at $t\approx40$ (Sec.~\ref{sec:setup.kerr}). No single Kerr metric matches the data over the whole window. The values fitted on $0.6\le t\le62$, $M=1.0543$ and $a=0.4871$ (dashed lines), lie within the range spanned by the horizon. To test whether the early, unsettled part of the evolution causes the offset, we retrain on $40\le t\le62$ only (shaded region), keeping the network, initialization, sampling, and optimizer unchanged. We find that the residual metric mismatch is substantially smaller, so the late-time data are much closer to a single Kerr metric. The learned values become $M=1.0562$ and $a=0.4859$ (dash-dotted lines), close to the late-time horizon values $M_{\rm H}=1.0566$ and $a_{\rm H}=0.4856$. 

\section{Spacetime Metrics in General Relativity}
\begin{widetext}
\label{app:metrics}
\subsection{Schwarzschild Metric}
\subsubsection{Spherical Polar coordinates}
\begin{equation}
\mathbf{g}^{SC}_{\mu\nu} =
\begin{pmatrix}
-(1 - \frac{2M}{r}) & 0 & 0 & 0 \\
0 & (1 - \frac{2M}{r})^{-1} & 0 & 0 \\
0 & 0 & r^{2} & 0 \\
0 & 0 & 0 & r^{2}\sin^{2}\theta
\end{pmatrix}
\end{equation}
\subsubsection{Ingoing Eddington-Finkelstein Coordinates}
\begin{equation}
  \mathbf{g}^{EF}_{\mu\nu} =
\begin{pmatrix}
-(1 - \frac{2M}{r}) & 1 & 0 & 0 \\
1 & 0 & 0 & 0 \\
0 & 0 & r^{2} & 0 \\
0 & 0 & 0 & r^{2}\sin^{2}\theta
\end{pmatrix}  
\end{equation}

\subsection{Kerr Metric}

\subsubsection{Boyer--Lindquist Coordinates}
\begin{equation}
\mathbf{g}^{BL}_{\mu\nu}= \begin{pmatrix}
-\left( 1 - \frac{2Mr}{\rho^2} \right) & 0 & 0 & -\frac{2Mar \sin^2\theta}{\rho^2} \\
0 & \frac{\rho^2}{\Delta} & 0 & 0 \\
0 & 0 & \rho^2 & 0 \\
-\frac{2Mar \sin^2\theta}{\rho^2} & 0 & 0 & \left( r^2 + a^2 + \frac{2Ma^2r \sin^2\theta}{\rho^2} \right) \sin^2\theta
\end{pmatrix}
\end{equation}
where,
\begin{equation}
\rho^2 = r^2 + a^2 \cos^2 \theta, \quad \Delta = r^2 - 2Mr + a^2
\end{equation}

\subsubsection{Ingoing-Kerr Coordinates}
\begin{equation}
\mathbf{g}^{IK}_{\mu\nu}= \begin{pmatrix}
-\left( 1 - \frac{2Mr}{\rho^2} \right) & 1 & 0 & -\frac{2Mar \sin^2\theta}{\rho^2} \\
1 & 0 & 0 & -a\sin^{2}{\theta} \\
0 & 0 & \rho^2 & 0 \\
 -\frac{2Mar \sin^2\theta}{\rho^2}&-a\sin^{2}{\theta}  & 0 & \left( r^2 + a^2 + \frac{2Ma^2r \sin^2\theta}{\rho^2} \right) \sin^2\theta
\end{pmatrix}
\end{equation}
where,
\begin{equation}
\rho^2 = r^2 + a^2 \cos^2 \theta
\end{equation}
\subsubsection{Cartesian Kerr--Schild Metric}
\begin{equation}
\mathbf{g}^{KS}_{\mu\nu} =
\begin{pmatrix}
-1 + 2H 
& 2H\,\dfrac{r x + a y}{\,r^{2} + a^{2}\,} 
& 2H\,\dfrac{r y - a x}{\,r^{2} + a^{2}\,} 
& 2H\,\dfrac{z}{\,r\,} 
\\[1.2em]
2H\,\dfrac{r x + a y}{\,r^{2} + a^{2}\,} 
& 1 + 2H\!\left(\dfrac{r x + a y}{\,r^{2} + a^{2}\,}\right)^{2} 
& 2H\,\dfrac{(r x + a y)(r y - a x)}{(r^{2} + a^{2})^{2}} 
& 2H\,\dfrac{(r x + a y) z}{\,r(r^{2}+a^{2})\,} 
\\[1.2em]
2H\,\dfrac{r y - a x}{\,r^{2} + a^{2}\,} 
& 2H\,\dfrac{(r x + a y)(r y - a x)}{(r^{2}+a^{2})^{2}} 
& 1 + 2H\!\left(\dfrac{r y - a x}{\,r^{2} + a^{2}\,}\right)^{2} 
& 2H\,\dfrac{(r y - a x) z}{\,r(r^{2}+a^{2})\,} 
\\[1.2em]
2H\,\dfrac{z}{\,r\,} 
& 2H\,\dfrac{(r x + a y) z}{\,r(r^{2}+a^{2})\,} 
& 2H\,\dfrac{(r y - a x) z}{\,r(r^{2}+a^{2})\,} 
& 1 + 2H\!\left(\dfrac{z}{\,r\,}\right)^{2}
\end{pmatrix}.
\end{equation}

                where:
                \[
                H = \frac{M r^3}{r^4 + a^2 z^2}, \quad \text{and} \quad r^4 - r^2(x^2 + y^2 + z^2 - a^2) - a^2 z^2 = 0.
                \]
\end{widetext}
\bibliography{apssamp}
\end{document}

%% file: acronyms.tex
\newacro{adm}[ADM]{Arnowitt-Deser-Misner}
\newacro{amr}[AMR]{adaptive mesh refinement}
\newacro{ai}[AI]{artifical intelligence}
\newacro{bbh}[BBH]{binary black hole}
\newacroplural{bbh}[BBHs]{binary black holes}
\newacro{bms}[BMS]{Bondi--Metzner--Sachs}
\newacro{bh}[BH]{black hole}
\newacroplural{bh}[BHs]{black holes}
\newacro{bhns}[BHNS]{black hole-neutron star}
\newacro{bhpt}[BHPT]{black hole perturbation theory}
\newacro{bns}[BNS]{binary neutron star}
\newacro{bf}[BF]{Bayes' factor}
\newacro{cbc}[CBC]{compact binary coalescence}
\newacro{ce}[CE]{Cosmic Explorer}
\newacro{cce}[CCE]{Cauchy characteristic extraction}
\newacro{cfl}[CFL]{Courant--Friedrichs--Lewy}
\newacro{da}[DA]{data analysis}
\newacro{et}[ET]{Einstein Telescope}
\newacro{efe}[EFE]{Einstein Field Equation}
\newacro{ef}[EF]{Eddington--Finkelstein}
\newacro{eob}[EOB]{Effective-One-Body}
\newacro{eom}[EOM]{equations of motion}
\newacro{fd}[FD]{frequency domain}
\newacro{fre}[FRE]{finite-radius extraction}
\newacro{fd}[FD]{finite differencing}
\newacro{fft}[FFT]{Fast Fourier transform}
\newacro{gw}[GW]{gravitational wave}
\newacro{ghg}[GHG]{generalised harmonic gauge}
\newacro{gh}[GH]{generalised harmonic}
\newacro{gr}[GR]{general relativity}
\newacro{grb}[GRB]{gamma-ray burst}
\newacro{grhd}[GRHD]{general-relativistic hydrodynamics}
\newacro{gwosc}[GWOSC]{Gravitational Wave Open Science Center}
\newacro{gwtc1}[GWTC-1]{the first gravitational-wave transients catalog}
\newacro{gsf}[GSF]{Gravitational Self Force}
\newacro{hm}[HM]{Higher mode}
\newacroplural{hm}[HMs]{Higher modes}
\newacro{hpc}[HPC]{high-performance computing}
\newacro{ifo}[IFO]{interferometer}
\newacro{imr}[IMR]{inspiral-merger-ringdown}
\newacro{im}[IMR]{inspiral-to-merger}
\newacro{kagra}[KAGRA]{Kamioka Gravitational Wave Detector}
\newacro{ligo}[LIGO]{Laser Interferometer Gravitational-Wave Observatory}
\newacro{lisa}[LISA]{Laser Interferometer Space Antenna}
\newacro{lr}[LR]{Light Ring}
\newacro{lso}[LSO]{Last Stable Orbit}
\newacro{lvc}[LVC]{LIGO-Virgo Collaboration}
\newacro{lvk}[LVK]{LIGO-Virgo-Kagra}
\newacro{lo}[LO]{leading order}
\newacro{ml}[ML]{machine learning}
\newacro{ns}[NS]{neutron star}
\newacroplural{ns}[NSs]{neutron stars}
\newacro{nr}[NR]{numerical relativity}
\newacro{nn}[NN]{neural network}
\newacro{nqc}[NQC]{next-to-quasicircular}
\newacro{nlo}[NLO]{next-to-leading order}
\newacro{nnlo}[NNLO]{next-to-next-to-leading order}
\newacro{n3lo}[N3LO]{next-to-next-to-next-to-leading order}
\newacro{n4lo}[N3LO]{next-to-next-to-next-to-next-to-leading order}
\newacro{ode}[ODE]{Ordinary Differential Equation}
\newacroplural{ode}[ODEs]{Ordinary Differential Equations}
\newacro{pde}[PDE]{partial differential equation}
\newacro{pn}[PN]{post-Newtonian}
\newacro{pm}[PM]{post-Minkowskian}
\newacro{pe}[PE]{parameter estimation}
\newacro{psd}[PSD]{power spectral density}
\newacroplural{psd}[PSD]{power spectral densities}
\newacro{pa}[PA]{post-adiabatic}
\newacro{qnm}[QNM]{quasi-normal mode}
\newacro{qc}[QC]{quasi-circular}
\newacro{snr}[SNR]{signal-to-noise ratio}
\newacro{spa}[SPA]{stationary-phase approximation}
\newacro{sxs}[SXS]{Simulating eXtreme Spacetimes}
\newacro{td}[TD]{time domain}
\newacro{ng}[NG]{Nect Generation}
\newacro{dcs}[dCS]{dynamical Chern-Simons}
\newacro{bl}[BL]{Boyer--Lindquist}
\newacro{bgr}[BGR]{beyond-\ac{gr}}
\newacro{spec}[SpEC]{Spectral Einstein Code}
\newacro{edgb}[EdGB]{Einstein-dilaton-Gauss-Bonnet}
\newacro{emri}[EMRI]{extreme mass-ratio inspiral}
\newacro{gwtc}[GWTC]{Gravitational-Wave Transient Catalog}
\newacro{far}[FAR]{false alarm rate}
\newacro{ci}[CI]{credible interval}
\newacro{ks}[KS]{Kerr--Schild}